\documentclass[usenatbib]{mn2e}
\pdfoutput=1
\usepackage{float}
\usepackage{graphicx}
\usepackage{amsmath}
\usepackage{placeins}
\usepackage{threeparttable}
\usepackage{mathptmx}
\usepackage{txfonts}
\usepackage[T1]{fontenc}
\usepackage{ae,aecompl}
\usepackage{placeins}
\usepackage{hyperref}
\usepackage{pdfpages}
\usepackage{etex}
\usepackage{tablefootnote}
\usepackage{longtable}
\hypersetup{colorlinks,
citecolor=blue
}

\begin{document}

\title[How ``cold" are the superthins?]
{How ``cold" are the stellar discs of superthin galaxies?}
\author[K.Aditya \& Arunima Banerjee]
       {K.Aditya$^{1}$\thanks{E-mail: kaditya@students.iisertirupati.ac.in} and 
        Arunima Banerjee$^{2}$\thanks{E-mail :  arunima@iisertirupati.ac.in} \\
\{$1$, $2$\}  Indian Institute of Science Education and Research, Tirupati 517507, India \\ 
} 
\maketitle

\begin{abstract}
Superthin galaxies are a class of bulgeless, low surface brightness galaxies with strikingly high values of planar-to-vertical axes ratio $\rm(b/a> 10 - 20)$,
possibly indicating the presence of an ultra-cold stellar disc. Using the multi-component galactic disc model of gravitationally-coupled stars and gas in the
force field of the dark matter halo as well as the stellar dynamical code AGAMA (Action-based Galaxy Modelling Architecture), we determine the vertical 
velocity dispersion of stars and gas as a function of galacto-centric radius for five superthin galaxies (UGC 7321, IC 5249, FGC 1540, IC2233 and UGC00711) 
using observed stellar and atomic hydrogen (HI) scale heights as constraints, using a Markov Chain Monte Carlo Method. We find that the central vertical velocity 
dispersion for the stellar disc in the optical band varies between $\sigma_{0s}$ $\sim$ $10.2 - 18.4$ $\rm{kms}^{-1}$  and falls off with an exponential scale length of 
$2.6$ to $3.2$ $R_{d}$ where $R_{d}$ is the exponential stellar disc scale length. Interestingly, in the 3.6 $\mu$m, the same, 
averaged over the two components of the stellar disc, varies between $5.9$ to $11.8$ $\rm{kms}^{-1}$, both of which confirm the
presence of "ultra-cold" stellar discs in superthin galaxies. Interestingly, the global median of the multi-component disc dynamical stability
parameter $Q_N$ of our sample superthins is found to be 5 $\pm$ 1.5, which higher than the global median value of 2.2 $\pm$ 0.6 for a sample 
of spiral galaxies.
\end{abstract}


\section{INTRODUCTION}
Superthin galaxies are a class of edge-on disc galaxies exhibiting extra-ordinarily high values of planar-to-vertical axes ratio $b/a$ $\sim$ 10-20, with no
discernible bulge component. They are generally characterized by low values of central $B$-band surface brightness $\mu_{B}$ $\sim$ 23-26 $\rm{mag arcsec}^{-2}$, low star formation rates $\sim$ 0.01 - 0.05 $M_{\odot}{\rm{yr}}^{-1}$, gas richness as indicated by high values of the ratio of the total neutral hydrogen (HI) mass to the total $B$-band luminosity ${M_{HI}}/{L_{B}} \sim$ 1 and dark matter dominance at all galacto-centric radii. Superthins are therefore classic examples of under-evolved systems and ideal test-beds of galaxy formation and evolution processes in the local universe (See \citet{Matthews1999} for a review).

The term \emph{superthin} was first introduced by \cite{goad1981spectroscopic} who did a spectroscopic study of four edge-on galaxies: 
UGC 7321, UGC 7170,UGC 9242 and UGC 4278 (IC 2233). Superthin galaxies were also studied as part of Flat Galaxy catalog (FGC) which was 
an optical survey of flat and bulgeless galaxies in the local universe; 1150 out of 4000 FGC galaxies were found to have $\rm b/a >10$ \citep{karachentsev1993flat}. 
Later superthin galaxies have also been studied as part of the optical study of flat galaxies  \citep{Kautsch2009} and, recently, very thin galaxies in 
the SDSS \citep{bizyaev2016very}. In addition, being rich in neutral hydrogen gas (HI), superthin galaxies have also been studied as of large HI surveys of edge-on disc galaxies. 
See, for example, \cite{giovanelli1997spectroscopy} and \cite{matthews2000h}.

The origin of a superthin stellar disc in these low surface brightness galaxies is still not well understood. The vertical scale height of the stellar disc in 
a disc galaxy is determined by a balance between the gradient of the stellar velocity dispersion in the vertical direction and the net vertical gravitational potential. 
Using their multi-component galactic disc model of gravitationally-coupled stars and gas in the force-field of the dark matter halo as constrained by the observed HI rotation
curve and HI scale height, \cite{banerjee2010dark} found that the superthin galaxy UGC7321 has a dense and compact dark matter halo i.e., $\frac{R_{c}}{R_{d}} \leq 2$ where 
$R_c$ is the core radius of the pseudo-isothermal dark matter halo and $R_d$ the exponential stellar disc scale length (See, also, \citet{o2010dark}). As a direct follow-up of 
this work, \cite{banerjee2013some} showed that the compact dark matter halo is responsible for the existence of superthin disc in UGC 7321. Using stellar photometry 
and HI 21cm radio-synthesis,  mass models of a few superthin galaxies were constructed using the observed HI rotation curve only: IC5249, IC2233 \citep{banerjee2016mass} and 
FGC1540 \citep{kurapati2018mass}. In all these cases, it was found that $\frac{R_{c}}{R_{d}} \leq 2$, possibly indicating superthin galaxies are characterized by dense and compact 
dark matter halos in general, which, in turn, may strongly regulate the structure and dynamics of the galactic disc. \cite{zasov1991thickness} showed that a massive dark matter 
halo was responsible for suppressing bending instabilities in superthin galaxies. Using the 2-component disc dynamical stability parameter $Q_{RW}$ proposed 
by \citet{romeo2011effective}, \cite{GargBan2018} showed that the dark matter halo is responsible for the dynamical stability against local, axis-symmetric perturbations 
in a general sample of low surface brightness galaxies. Alternatively, given the fact that the morphology of disc galaxies is primarily driven by the angular momentum 
of their discs, the large planar-to-vertical axes ratios of the stellar discs in superthin galaxies may possibly be the outcome of a relatively higher value of the specific 
angular momentum of their discs. \cite{JadhavBan2019} however found that within the 95.4$\%$ confidence interval, some of the superthins does obey the same angular 
momentum-mass relation as ordinary disc galaxies, thus ruling out the role of the specific angular momentum as the primary factor in regulating the vertical structure
of the superthin discs.

Finally, the origin of the superthin stellar disc may be possibly linked with small values of the vertical stellar velocity dispersion, which is indicative of minimal disc 
heating in a direction perpendicular to the galactic plane. Recent advances in Integral Field Unit (IFU) astronomy surveys  have successfully estimated well-resolved stellar
velocity dispersion for face-on or nearly face-on galaxies (\citet{Capp2011},  \citet{law2015observing}, \citet{allen2015sami}, \citet{Bershady_2010}, \citet{sanchez2012califa}). 
However, due to the edge-on geometry of the superthin galaxies, the direct determination of the vertical velocity dispersion is not feasible. 
In this paper, we use the multi-component galactic disc model of gravitationally-coupled stars and gas in the force field of the dark matter halo 
to constrain the vertical stellar dispersion for five superthin galaxies: UGC 7321, IC 5249, FGC 1540, IC2233 and UGC711 in the optical as well as the 3.6 $\mu$m using 
observed scale height data as constraint and employing the Markov Chain Monte Carlo (MCMC) method \citep{narayan2002vertical}. The mass models for the above galaxies constructed 
using stellar photometry and HI radio-synthesis observations were already available in the literature. Further, we check the consistency of our results from the multi-component 
model by using the publicly available stellar dynamical code Action-based Galaxy Modelling Architecture (AGAMA) of \citet{vasiliev2018agama}. We use the best-fit stellar dispersion
from the multi-component model as an input to AGAMA and determine the stellar vertical scale heights. We then compare it with the observed stellar scale height used to constrain the
self-consistent model to check for the robustness of our results. Finally, we check the dynamical stability of our model galactic discs by calculating the multi-component disc 
stability parameters as proposed by \citet{romeo2011effective} and \citet{romeo2013simple}.

The paper is organized as follows: In \S 2 we introduce the dynamical models of the galaxy and the dynamical stability parameters of the multi-component galactic discs,
in \S 3, we describe the basic structural properties of our sample superthin galaxies and in the \S 4 the observational constraints for our each sample galaxy. 
In \S 5, we present our results and the corresponding discussion, followed by conclusions in \S 6.

\section{Dynamical model of the galaxy}
\subsection{The multi-component model}
We model the galaxy as a multi-component system of gravitationally-coupled stars and HI gas, in the external force field of a dark matter halo. 
We further assume that the stars and gas are in vertical hydrostatic equilibrium and that their velocity dispersions remain constant in the $z$-direction. 
Finally, for reasons of simplicity, we assume that the stars and gas are confined in axisymmetric discs, which are coplanar and concentric with each other.
The joint Poisson distribution for the above system in terms of galactic cylindrical coordinates $(R,\phi,z)$ is ;
\begin{equation}
\frac{\partial ^{2} \Phi _{\rm{total}}}{\partial z^{2}} +\frac{1}{R} \frac{\partial}{\partial R}(\frac{R \partial \Phi _{\rm{total}}}{\partial R}) =
4 \pi G(\sum_{i=1}^{n}\rho _{i} + \rho_{DM})
\end{equation}
where $\Phi _{\rm{total}}$ is the total gravitational potential due to the disc components and the dark matter halo and $\rho _{i}$ is 
the density of the i$^{\rm{th}}$ disc component where $i = 1$ to $n$, $n$ denoting the number of disc components. $\rho_{DM}$ the density of the dark matter halo.
For a galaxy with a flat rotation curve, the radial term drops out and the Poisson's equation reduces to
\begin{equation}
\frac{\partial ^{2} \Phi _{\rm{total}}}{\partial z^{2}}  =
4 \pi G(\sum_{i=1}^{n}\rho _{i} + \rho_{DM})
\end{equation}

The equation of vertical hydrostatic equilibrium for the $j^{th}$ component of the disc $(j=stars,gas)$ \citep{rohlfs1977lectures} is 

\begin{equation}
\frac{\langle(\sigma_{z})^{2}_{j}\rangle}{\rho_{j}}\frac{\partial \rho_{j}}{\partial {z}} + \frac{\partial \Phi _{\rm{total}} } {\partial z} =0
\end{equation} 

Combining the joint Poisson's equation and the equation for vertical hydrostatic equilibrium we get:
\begin{equation}
\frac{\partial^{2}\rho_{j}}{\partial z^{2}} = \sum_{i=1}^{n} -4\pi G
\frac{\rho_{j}}{\langle(\sigma_z)_{j}^2\rangle} (\rho_{i} +  \rho_{DM}) +
(\frac{\partial\rho_j}{\partial z})^2 \frac{1}{\rho_{j}};
\end{equation}
where $\rho_{j}$, $j$ stands for the density and $\langle(\sigma_z)_{j}^2\rangle$ is the vertical velocity dispersion of the $j^{th}$ component with $j = 1$ to $n$.
The dark matter is modelled as a pseudo-isothermal profile given by
\begin{equation}
\rho_{DM}=\frac{\rho_{0}}{(1+\frac{m^{2}}{R^{2}_{c}})}
\end{equation}
where     
\begin{equation}
m^2= R^2+\frac{z^2}{q^2}
\end{equation}
\citep{de1988potential}
where $\rho_0$ is the central core density, $R_c$ the core radius and $q$ the vertical-to-planar axes ratio of the spheroidal the halo. For a spherical halo 
$q=1$, oblate $q<1$, prolate $q>1$. We assume a spherical halo in this work. In our work, $\rho_{DM}$ is an input parameter, which was already determined in 
earlier mass modelling studies.
The radial profile of the vertical velocity dispersion of each of the stellar disc components is parametrized as :
\begin{equation}
\sigma_{z,s}(R)=\sigma_{0s} \exp(-R/{\alpha_{s}}R_{d})
\end{equation} 
\noindent Here $\sigma_{0s}$ is the central value of the vertical velocity dispersion of the stars and $\alpha_{s}$ the radial scale length of the 
exponential fall-off of the same in units of the exponential disc scale length $R_d$. Both $\sigma_{0s}$ and $\alpha_{s}$ are free parameters. This is closely 
following the work of \cite{van1981}, who modelled galactic discs of a sample of edge-on disc galaxies as self-gravitating with a vertical velocity dispersion
remaining constant with $z$, and found $\alpha_{s} = 2$. However, low surface brightness galaxies like the superthins are gas-rich as well as dark matter dominated and 
hence cannot be modelled as self-gravitating discs. Hence, although we adopt the above parametric form for the radial profile of the vertical velocity dispersion of the 
stars for our sample superthins, we keep $\alpha_{s}$ as a free parameter in our model.
Finally, the radial profile of the HI vertical velocity dispersion is parametrized as a polynomial as follows:
\begin{equation}
\sigma_{z,HI}(R) = \sigma_{\rm{0HI}} + \alpha_{HI} R +\beta_{HI} R^{2} 
\end{equation}
with $\sigma_{\rm{0HI}}$, $\alpha_{HI}$ and $\beta_{HI}$ as free parameters. This is similar to the parametrizations adopted in
modelling the HI velocity dispersion in M31 \citep{nar2005}  and in the Milky Way \citep{ban2008}. In some cases, the above profile may give a 
bad fit to the observed data and therefore we had to use a different profile as given below in order to get a better fit with the observed data.
\begin{equation}
\sigma_{z,HI}(R)=\sigma_{0HI}e^{-R/{\alpha_{HI}}}
\end{equation}
with $\sigma_{0HI}$ and $\alpha_{HI}$ as free parameters
Equation (4) thus represents $n$ coupled, non-linear ordinary differential equations in the variables $\rho_{\rm{i}}$ where i = 1 to $n$.  
For a given set of values of the free parameters, the above equation determines $\rho_{i}$'s as a function of $z$ and hence scaleheight for all $i$ at any $R$. 
The parameter values and hence the velocity dispersion profiles are constrained by the observed scale height values.
The above equation is solved iteratively using the Runge-Kutta method with boundary conditions at midplane $z=0$ given by;
\begin{equation}
\frac{d\rho_i}{dz}
\end{equation}
and    
\begin{equation}
\rho_{i}=(\rho_{0})_{i}    
\end{equation}
However, ${\rho}_i$ at $z=0$ is not known a priori. Instead the surface density ${\Sigma}_i$(R) (See equations 24 to 26 in section 4), which is given by twice the area under curve of ${\rho}_i$(z) versus
$z$, is used as the second boundary condition, since ${\Sigma}_i$(R) can be observationally determined. Hence the required value of ${\rho}_i$(0) can be then fixed by trial 
and error method, which eventually dtermines the ${\rho}_i$(z) distribution.    
The above method has been used to study vertical density distribution of stars and gas in a host of disc galaxies as in \cite{narayan2002origin},
\cite{nar2005},\citet{banerjee2007origin}, \citet{ban2008},\citet{banerjee2010dark},  \citet{ban2011a},\citet{ban2011b}, \citet{banerjee2013some}, \cite{sarkar2019flaring}. \\
      
\noindent \textbf{Model Fitting using Markov Chain Monte Carlo Method}: Since the parameter space to be scanned to obtain the best-fitting model is 
higher ($4 - 7$) dimensional (\S 4), we use the Markov Chain Monte Carlo (MCMC) method for determining the best-fitting set of parameters of our model. 
We use the task modMCMC from the publicly available R package FME \citep{soetaert2010inverse}, which implements MCMC using adaptive Metropolis procedure 
\citep{haario2006dram} \\

\noindent \textbf{Mean vertical stellar velocity dispersion:} As we will see in \S 4, a stellar disc may be represented as a superposition of 
two exponential discs in a given photometric band. We will represent the mean velocity dispersion of such a stellar disc by introducing the density 
averaged mean dispersion given by:
\begin{equation}
\sigma^{2}_{z,s(avg)}= \frac{\rho_{1}\sigma^{2}_{z1} +\rho_{2}\sigma^{2}_{z2} }{\rho_{1} + \rho_{2} }
\end{equation}
\noindent the subscripts $1$ and $2$ denoting disc 1 and disc 2 respectively. 

\subsection{AGAMA}
We use the publicly available stellar dynamical code AGAMA by \citet{vasiliev2018agama} 
\footnote{https://github.com/GalacticDynamics-Oxford/Agama}  for an alternative dynamical modelling our sample of galaxies. Here we model each stellar disc at a time,
assuming that it responds to the composite gravitational potential of all the disc components and the dark matter halo. We assume a double exponential profile for the 
same i.e. with exponential density distribution in the radial as well as in the vertical direction. We assume the HI component as having an exponential radial surface density 
profile with a constant scale height. The dark matter halo is modelled to have a  pseudo-isothermal density profile. We then bind together all the dynamical components to 
construct a composite potential. We then initialize a quasi-isothermal distribution function for the stellar disc with the composite potential and the requisite structural 
parameters of the stellar disc, including the stellar velocity dispersion as obtained from our multi-component galactic disc model. The distribution function and the composite 
potential are then combined with an action finder for constructing a galaxy model with a single stellar population responding to the net underlying gravitational potential of 
the galaxy. The quasi-isothermal distribution function (DF) is given by
\begin{equation}
 f(J)=f_{0}(J_{\phi}) \frac{\kappa}{\sigma_{R,s}^{2}} e^{- \frac{-\kappa J_{R}}{\sigma_{R,s}^{2}}}\frac{\nu}{\sigma_{z,s}^{2}}e^{\frac{-\nu J_{z}}{\sigma_{z}^{2}}}
\end{equation}
where $\kappa$ and $\nu$ are the radial and vertical epicyclic frequencies respectively. $\sigma_{R,s}$ and $\sigma_{z,s}$ are the stellar velocity dispersions in the 
$R$ and $z$ directions respectively. $J_{R}$, $J_{z}$ and $J_{\phi}$ are the actions of the stellar discs in the $R$, $z$ and $\phi$ directions respectively. Here 
$J^{2}_{\phi}=R^{3} \frac{\partial \Phi }{\partial R}$ and $J^2 = {J_R}^2 + {J_{\phi}}^2 + {J_{z}}^2$. \\
Moments of the density function may be computed as follows:
The stellar density is given by
\begin{equation}
{\rho}_s(x)= \int \int \int d^{3}v f(J[x,v ])
\end{equation}
The mean stellar velocity is given by
\begin{equation}
\bar{v}=   \frac{1}{{\rho}_s} \int \int \int d^{3}v v f(J)
\end{equation}
while the second moment of stellar velocity is given by
\begin{equation}
 \overline{v_{ij}^{2}} =\frac{1}{\rho} \int \int \int d^{3}v v_{i} v_{j} f(J)
\end{equation}
 Hence the velocity dispersion tensor is defined as
 \begin{equation}
\sigma_{s, ij}^{2} =\overline{v_{ij}^{2}} - \overline{v_{i}}\overline{v_{j}}
\end{equation}
Each stellar disc is modeled using a 'Disk' type potential. The density distribution due to the disk type potential is given by
\begin{equation}
\rho_s = \Sigma_{0s}\,\exp\big(-\big[\frac{R}{R_\mathrm{d}}\big]^{\frac1n} - \frac{R_\mathrm{cut}}{R}\big)
\times\left\{ \begin{array}{ll} \delta(z)\qquad\qquad\mbox{if} & h_z = 0 \\[1mm]
\frac{1}{2h_z} \exp\big(-\big|\frac{z}{h_z}\big|\big) & h_z > 0 \\[1mm]
\frac{1}{4|h_z|}\, \mathrm{sech}^2\big(\big|\frac{z}{2h_z}\big|\big) & h_z < 0 \end{array} \right. 
\end{equation}
\noindent where $\Sigma_{0s}$ is the central surface brightness, $R_{d}$ is the disc scale length, $h_z$ is the disc scale height, $R_{cut}$ is the disc inner cut-off and $n$ the sersic index. 
The HI component is modeled with a 'Disk' type density profile, with sersic index $n=0.5$ and using the average HI scaleheight. 
The dark matter density is modeled using 'Spheroid' type potential  with $\alpha =2$, $\gamma=0$, and $\beta =2$ where 'spheroid'type density is given by
\begin{equation}
\rho = \rho_0  \left(\frac{\tilde r}{a}\right)^{-\gamma} \Big[ 1 + \big(\frac{\tilde r}{a}\big)^\alpha \Big]^{\frac{\gamma-\beta}{\alpha}}\times \exp\Big[ -\big(\frac{\tilde r}{r_\mathrm{cut}}\big)^\xi\Big]
\end{equation}
\noindent where $\rho_{0}$ is the central density and $a$ the core radius.
We add together the above densities to create a total density profile of the galaxy. Then we use the 'GalaxyModel' function to 
create a composite model of the galaxy using the total density profile and the quasi-isothermal distribution for the disc component. 
Using the tasks 'moments' and 'projectedMoments' we compute the radial and vertical stellar dispersion profiles and the scale height of the stellar disc. We note here 
we do not employ the iterative method for constructing self-consistent equilibrium configurations, but simply initialize a DF in the given composite potential of all 
the disc components and the dark halo. \\
\noindent \textbf{Multi-component galactic disc model versus AGAMA model:} We note here, we are comparing two approaches for computing 
the stellar vertical dispersion: one is based on the Jeans' equations for hydrostatic equilibrium, generalized to include multiple components 
i.e, both stars and gas, and the other on computing the moments of a distribution function in the given potential. Both methods are based on the equilibrium
assumption, but differ in details. The common limitation of the former method is the neglect of radial gradients i.e. the resulting ordinary differential equation 
is solved in $z$ direction independently at each $R$, and the latter method is in principle more general and accurate, but as long as the potential is indeed 
computed self-consistently from the DF. In that case, the full set of Jeans equations (and not just the vertical one) should be satisfied automatically. However, the caveat 
is the density profile generated by the DF does not necessarily follow the exponential law, although it should be reasonably close if the parameters 
of the DF were chosen correctly, and if the system is not too hot. The other important parameters of the DF are the central value of radial velocity dispersion, 
and the scale length of its fall off which cannot be obtained from the multi-component model directly, in a self-consistent manner. We note here we do not employ 
the iterative method for constructing self-consistent equilibrium configurations, but simply initialize a DF in the given composite potential of all the disc components 
and the dark halo. This approach will give reasonable results, provided the parameters of the DF are in agreement with the parameters of the stellar density
profile (i.e., central surface density, scale radius, scale height are the same in the Disk density profile and in the QuasiIsothermal DF) which has been ensured in this work.

\subsection{Disc dynamical stability}
The disc stability against local axis-symmetric perturbations is determined by the a balance between the 
self-gravity on one hand and combined effect of velocity dispersion and the centrifugal force due to coriolis spin up of the perturbations on the other hand. The $Q$ parameter \citep{toomre1964gravitational} for a one component rotating fluid discs is 
\begin{equation}
Q= \frac{\kappa \sigma_{R} }{\pi G \Sigma}
\end{equation}
where $\kappa$ is the epicyclic frequency given by $\kappa^{2}=-4B\Omega$. $B$ and $\Omega$ are the Oort constant and the angular frequency respectively,    
$\sigma_{R}$ is the radial velocity dispersion and $\Sigma$ the surface density at a given radius $R$. A value of $Q > 1$ implies a stable disc and $Q \leq 1 $ is indicative
of an unstable galactic disc which is characteristic of star-forming regions. Superthin galaxies are rich in gas, which may strongly regulate the disc dynamics closer to the midplane \citep{banerjee2007origin}. 
Hence, the galactic disc can no more be considered as a single component self-gravitating disc. We use multi component stability parameter $Q_{N}$ derived by \cite{romeo2013simple} for studying the stability of the galaxy disc in $3.6 \mu m$ where the galaxies are 
composed of two stellar discs and a HI, discs. The effective stability parameter $Q_{N}$ for a multi-component galaxy disc is defined as
\begin{equation}
\frac{1}{Q_{N}} =\sum_{i=1}^{N} \frac{W_{i}}{T_{i}Q_{i}}
\end{equation}
The thickness of the galaxy disc increases the effective stability of the galaxy disc and is parameterized as 
\begin{equation}
 T \approx \left\{
                \begin{array}{ll}
                  1+0.6(\frac{\sigma_{z}}{\sigma_{R}}) ^{2} \hspace{0.5cm}   if  \hspace{0.5cm}  0 \leq \frac{\sigma_{z}}{\sigma_{R}} \leq 0.5    \\       
                  0.8+0.7(\frac{\sigma_{z}}{\sigma_{R}})  \hspace{0.5cm}   if  \hspace{0.5cm}  0.5 \leq \frac{\sigma_{z}}{\sigma_{R}} \leq 1           
                 \end{array}
                 \right.
\end{equation}
The weight factor for the $Q_{N}$ is  defined as 
\begin{equation}
W_{i} =\frac{2\sigma_{m}\sigma_{i}}{\sigma_{m}^{2} + \sigma_{i}^{2}}
\end{equation}
Where $i$ is the $i^{th}$ component of the galaxy and m is the component with smallest $TQ$ i.e $T_{m}Q_{m} =min(T_{i}Q_{i})$.
We note that $Q_{RW}$ = 1 or $Q_N$ = 1 gives the critical stability level of a galactic disc in the presence of local, axi-symmetric perturbations only. 
However, real galactic discs are not subjected to local, axisymmetric perturbations alone. In fact, non-axisymmetric perturbations are primarily 
responsible for the formation of bars and spiral arms in galactic discs. \citet{Griv2012} showed that non-axisymmetric perturbations have 
a destabilizing effect and therefore may increase the critical stability threshold for local axisymmetric perturbations 
to $Q_{RW}$ $\sim$ 1 - 2 \citep{10.1093/mnras/stv1220}. In addition to this, there may be gas dissipation effects, which may raise the critical 
stability level further to $Q_{RW}$ $\sim$ 2-3 \citep{Elm2011}. Therefore, in this work, we consider a galactic disc to be dynamically stable 
if $Q_{RW}$ or $Q_N$ is between 2 and 3.

\section{Sample of superthin galaxies}
In order to constrain the vertical velocity dispersion, we need a sample of galaxies with observed stellar and HI scaleheight. As input parameters to the two-component model 
and for calculating the potential for setting up an equilibrium distribution function in AGAMA, we need the stellar and HI densities and along with the dark matter density
as a function of the galactocentric radius. Thus, we have chosen our sample of superthin galaxies based on the availability of the three dimensional stellar photometry in optical 
and 3.6 $\mu$m band along HI surface densities and mass models derived from high resolution HI 21 cm line observations.
\begin{table*}
\begin{center}
\hfill{}
\caption{Input parameters for the model arising from the observational constraints}
\begin{tabular}{|l|c|c|c|c|c|c|c|c|c|c|c|c|}
\hline
Parameters& $\mu^{\textcolor{red}{1}}_{01}$ &$\Sigma^{\textcolor{red}{2}}_{01}$ & $R^{\textcolor{red}{3}}_{d1}$& $h^{\textcolor{red}{4}}_{z1}$ &$\mu^{\textcolor{red}{5}}_{02}$ &$\Sigma^{\textcolor{red}{6}}_{02}$ & $R^{\textcolor{red}{7}}_{d2}$&$h^{\textcolor{red}{8}}_{z2}$ &$\rho^{\textcolor{red}{9}}_{0}$&$R^{\textcolor{red}{10}}_{c}$   \\
Galaxy          &$mag$ $arcsec^{-2}$&$M_{\odot}$ $pc^{-2}$&kpc&kpc&$mag$ $arcsec^{-2}$&$M_{\odot}$ $pc^{-2}$&kpc&kpc&$M_{\odot}$ $pc^{-3}$&kpc \\
\hline    
\hline
Structural properties in optical band\\
\hline
UGC 7321 &-&-&-&-&  23.5 &34.7 & 2.1 & 0.150 &0.039&2.99 \\

FGC 1540 &21.67&33.14&1.29&0.675&20.60&88.79&1.29&0.185&0.308&0.64\\

IC  2233 &  22.90&17.85&2.47&0.332&-&-&-&-&0.0457&1.84\\
UGC 00711&    -   &15.0&1.6&0.317 &-&-&-&-&0.05&2.9 \\
\hline
Structural properties in $3.6\mu$m \\
\hline
UGC 7321 &21.73&7.165&2.39&0.436&19.9&37.26&1.0&0.134&0.140&1.27\\
IC  5249 & 21.7&5.44&5.24&0.724&20.53&15.97&1.23&0.253&0.026&2.99\\
FGC 1540 &22.23&3.37&1.85&0.43&21.39& 8.167&0.54&0.152&0.319&0.63\\
IC  2233 &21.67&5.59&2.16&0.39&20.53&12.2&0.81&0.08&0.055&1.83\\
UGC 00711& -   &14.6&2.14&0.44&- &-&-&-& 0.033&2.95 \\
\hline
Parameters of HI disc\\
\hline
Parameters&$\Sigma^{HI\textcolor{red}{11}}_{01}$& $R^{\textcolor{red}{12}}_{0,1}$&$a^{\textcolor{red}{13}}_{1}$&$\Sigma^{HI \textcolor{red}{14}}_{02}$& $R^{\textcolor{red}{15}}_{0,2}$&$a^{\textcolor{red}{16}}_{2}$ \\   
          &$M_{\odot}$ $pc^{-2}$& kpc    &   kpc &$M_{\odot}$ $pc^{-2}$& kpc    &   kpc\\
\hline
 UGC 7321&  4.912&2.85&3.85&2.50&1.51&0.485\\
 IC  5249 &  3.669&3.35&5.92& 4.85&4.05&17.06\\
 FGC 1540&   4.09&5.73&2.48&1.3&5.73&5.08\\
 IC  2233 &  2.236&1.79&2.52&2.454&1.69&6.14\\ 
 UGC 00711 & 30.83&3.73&    &     &   &     \\
 \hline
\end{tabular}

\end{center}

\begin{tablenotes}
\item ($\textcolor{red}1$):  Central surface brightness of disc(1)

\item ($\textcolor{red}{2}$):  Central stellar surface density disc(1)\\
\item ($\textcolor{red}{3}$):  Exponential scale length for disc(1)\\
\item ($\textcolor{red}{4}$):  Exponential scale height for disc(1)\\
\item ($\textcolor{red}{5}$):  Central surface brightness of disc(2) \\
\item ($\textcolor{red}{6}$):  Central stellar surface density disc(2)\\
\item ($\textcolor{red}{7}$):  Exponential scale length for disc(2)\\
\item ($\textcolor{red}{8}$):  Exponential scale height for disc(2)\\
\item ($\textcolor{red}{9}$):  Core density of the pseudo-isothermal dark matter halo\\
\item ($\textcolor{red}{10}$): Core radius of the pseudo-isothermal dark matter halo\\
\item ($\textcolor{red}{11}$): The central surface density of disc 1 constituting the double gaussian HI profile\\
\item ($\textcolor{red}{12}$): Scalelength of gaussian disc(1)\\ 
\item ($\textcolor{red}{13}$): Offset in the centre of disc(1)\\
\item ($\textcolor{red}{14}$): The central surface density of disc 2 constituting the double gaussian HI profile\\
\item ($\textcolor{red}{15}$): Scalelength of gaussian disc(2)\\ 
\item ($\textcolor{red}{16}$): Offset in the centre of disc(2)\\
\end{tablenotes}
\end{table*}

\subsection{UGC7321}    
UGC 7321 is a protypical nearby superthin galaxy at a distance $D=10$ Mpc \citep{matthews2000extraordinary}, inclination $i=88^{\circ}$ \citep{matthews1999extraordinary} 
and major-to-minor axes ratio $b/a=10.3$. Its characterized by a steeply rising rotation curve with an asymptotic velocity $\sim$ 110 kms$^{-1}$\citep{uson2003hi}. 
The deprojected central surface brightness in B-band is 23.5 mag arcsec$^{-2}$\citep{matthews1999extraordinary}.
The galaxy has large dynamical mass  $M_{dyn}/M_{HI}=31$ and $M_{dyn}/L_{B}=29$ \citep{roberts1994physical} which underscores the importance of the dark matter in these galaxies.
Constraining the dark matter halo of UGC7321 using observed HI rotation curve and HI vertical scaleheight data revealed a compact dark matter with 
$\rho_{0}=0.039  M{\odot} pc^{-3}$ and $R_{c}=2.9kpc$ \citep{banerjee2010dark}. See, also, 
\citet{banerjee2016mass}.

\subsection{IC 5249}
IC 5249 is an edge-on galaxy observed at an inclination $i=89^{\circ}$ \citep{abe1999observation} with axial ratio $b/a =10.2 $. The 
galaxy has an asymptotic rotational velocity of about 112 kms$^{-1}$. The
${M_{dyn}}/{M_{HI}}=9.5$ and  $M_{dyn}/L_{B}=9.5$ (\citet{yock1999observation}, \citet{van2001kinematics}). Mass modelling of IC 5249  
indicated the presence of a dark matter halo with $R_{c}=2.9 kpc$ and $\rho_{0}=0.026 M_{\odot} pc^{-3}$ \citep{banerjee2016mass}.
\subsection{FGC 1540}
FGC 1540 is a superthin galaxy at a distance of $D=10$ Mpc. It is observed at an inclination of $i=87^{\circ}$ and has an
axial ratio $b/a = 7.5$. It is classified as a superthin in the Flat galaxy catalogue \citep{karachentsev1993flat}.
It has $M_{HI}/L_{B}=4.1$ and is characterised by an asymptotic rotational velocity of about 90 kms$^{-1}$ and the mass modelling 
indicates a central dark matter density $\rho_{0}=0.262 M_{\odot}/ pc^{3}$ and a core radius $0.69$ kpc
 \citep{kurapati2018mass}.
 
\subsection{IC 2233}
IC 2233 is a superthin galaxy with an axial ratio  b/a = 7,  observed at an inclination of $88.5^{\circ}$ \citep{matthews2007h} at a distance of 10 Mpc.
The galaxy has an aymptotic rotational velocity of about 85 kms$^{-1}$. $M_{HI}/L_{b}\sim0.62$ and the $M_{dyn}/M_{HI} \sim 12$, the latter indicating that the galaxy is 
rich in HI. Mass models predicts a central dark matter density  $\rho_{0}=0.055 M_{\odot} pc^{-3}$ and a 
core radius $1.83 kpc$ \citep{banerjee2016mass}. 

\subsection{UGC00711}
UGC0711 is a superthin galaxy with a planar-to-vertical axial ratio b/a = 15.5 and observed at an inclination of $74.7^{\circ}$ at a distance of $D = 23.4$ Mpc.
The galaxy has an aymptotic rotational velocity of $\sim$ 100 kms$^{-1}$ \citep{mendelowitz2000rotation}. Mass modelling  predicts a central dark matter density  
$\rho_{0}=0.033 M_{\odot} pc^{-3}$ and a charactaristic core radius $2.95 kpc$ \citep{banerjee2016mass}. 

\section{Observational constraints}
We model the vertical stellar dispersion of the superthin galaxies in optical and in the 3.6 $\mu$m band using observed stellar and HI scaleheight as a constraint.
The stellar disc appears superthin in optical, which, in turn, traces the young stellar population. The 3.6 $\mu$m band, on the other hand, traces a relatively 
older stellar population which also constitutes the major mass fraction of the stellar component. More importantly, is free from dust extinction. 
Except for FGC1540, all our sample superthins are characterized by a single exponential stellar disc in the optical. Similarly, in the 3.6 $\mu$m band, except for UGC711,
our sample galaxies are found to consist of  two exponential stellar discs. Therefore, for our sample stellar discs, the surface density is either a single exponential given by
\begin{equation}
{\Sigma}_s(R) = {\Sigma}_{0s} \rm {exp} (-R/R_d) 
\end{equation}
where ${\Sigma}_{s0}$ is the central stellar surface density and $R_d$ the exponential stellar disc scale length.
or, a double exponential given by
\begin{equation}
{\Sigma}_s(R) = {\Sigma}_{01} \rm{exp} (-R/R_{d1})) + {\Sigma}_{02} exp(-R/R_{d2}) 
\end{equation}
where ${\Sigma}_{01}$ is the central stellar surface density and $R_{d1}$ the exponential stellar disc scale length of stellar disc 1 and so on. However, we may note here a recent study of UGC7321 showed that a double disc is not 
required to explain the data and argued that physically the existence of a second, thick disk in superthin galaxies is debatable \citep{sarkar2019flaring}. \\
The structural parameters for the stellar disc, i.e., the central surface density, the disc scale length and the scale height for UGC 7321 in $B$-band were taken 
from \cite{uson2003hi}. For IC5249, structural parameters for the disc were not available in the literature. For FGC1540, the $i$-band parameters were taken 
from \cite{kurapati2018mass}. For IC2233, the same in $r$-band were obtained from \cite{bizyaev2016very}. Finally, the data for UGC711 in $B$-band were taken 
from \cite{mendelowitz2000rotation}. The structural parameters of the stellar discs of our sample galaxies in the optical band are summarized in Table 1. 
In 3.6 $\mu$m band, all our sample galaxies were found to have a thick and thin stellar disc, each with an exponential surface density and constant scaleheight.
The structural parameters for the same were taken from \cite{salo2015} and are presented in Table 1. To sum up, among our sample galaxies, only UGC7321, UGC00711, and IC2233 are 
characterised by a single exponential disc in the optical band, but a pair of exponential discs in the 3.6 $\mu$ band. It is interesting to observe that the scale lengths of 
the optical discs of the above galaxies are closer to those of the thick disc component in the 3.6 $\mu$m band. However, their central surface densities and 
vertical scale heights in the optical band are closer to those of the thin disc component in the 3.6 $\mu$ band. This already indicates that there is perhaps
no direct correspondence between the optical disc and any of the components of the 3.6 $\mu$m stellar disc for the above galaxies. FGC 1540 is characterised by
a pair of exponential discs in both the optical and the 3.6 $\mu$m band. The radial scale lengths of the two discs in the optical have the same value, and are close
to the value of that of the thick disc in the 3.6 $\mu$m band. However, there is a significant mismatch between the surface density and the scale height values of the
components in the optical and the 3.6 $\mu$m band. Finally, UGC711 is characterised by a single exponential disc both in the optical as well as in the 3.6 $\mu$m band, and 
the parameters seem to be fairly comparable to each other, possibly indicating that the disc is one and the same only in this case.

For UGC 7321, the HI surface density  was taken from \cite{uson2003hi}, for IC 5249 from \cite{van2001kinematics}, for FGC 1540 from \cite{kurapati2018mass}, 
for IC2233 from \cite{matthews2007h} and for UGC711 \cite{mendelowitz2000rotation}.  Earlier work indicated that the radial profiles of HI surface density could be
well-fitted with double-gaussians profiles (See, for example, \cite{patra2014modelling}), possibly signifying the presence of two HI discs. Also,
galaxies with the HI surface density peaking away from the centre are common, which indicates the presence of an HI hole at the centre. Our sample HI
surface density profiles could therefore be fitted well with off-centred double Gaussians given by
\begin{equation}
{\Sigma}_{HI} (R) = {\Sigma}^{HI}_{01} \rm{exp} [-{\frac{{(R - a_1)}^2}{2 {R_{0,1}}^2}}] + {\Sigma}^{HI}_{02} \rm{exp} [-{\frac{{(R - a_2)}^2}{2 {R_{0,2}}^2}}] 
\end{equation}
where ${\Sigma}^{HI}_{01}$ is the central gas surface density, $a_1$ the centre and $R_{0,1}$ the scale length of gas disc 1 and so on. 
For the gas disc, we consider the atomic hydrogen (HI) surface density only as the presence of molecular gas in LSBs is known to be negligible 
(See, for example, \cite{banerjee2016mass} for a discussion). The HI scaleheight for UGC 7321 and IC 5249 were obtained from \cite{o2010dark}. 
For FGC 1540, we used approximately constant HI scaleheight of 0.400 kpc (Kurapati, private communication). HI scaleheigth for the galaxy IC2232 and UGC711 was obtained 
by the using the FWHM vs $\frac{2R}{D_{HI}}$ plot as given in \cite{o2010dark} as a scaling relation. We obtain FWHM= $\frac{2.4}{0.5D_{HI}}R +0.244$, where $D_{HI}$ is 
the HI diameter. We note here that due to the unavailability of observed HI scaleheight data, we have used rough estimates of the same in case of FGC1540, IC2233 and UGC711 
in our calculations as discussed above. However, we stress that the the HI scaleheight tightly constrains the value of HI velocity dispersion and not so much that of 
the stellar disc. So reasonable variation in the assumed value of the HI scaleheight will hardly change the best-fitting values of stellar vertical velocity dispersion as 
determined by our model. The parameters of the HI disc are summarized in Table 1.

For the stellar disc modelled using the optical band, the dark matter profile parameters i.e central core density $\rho_{0}$ and the core radius $R_{c}$ for UGC 7321, 
was modelled by constructing mass models using the 'rotmas' and 'rotmod' tasks in gipsy \citep{van1992groningen}. The same for FGC1540 were taken from \cite{kurapati2018mass}. 
For IC2233 and UGC711, the same were similarly modelled using gipsy. For the stellar disc modelled using the 3.6 $\mu$m band, the dark matter profile parameters i.e central 
core density $\rho_{0}$ and the core radius $R_{c}$ for UGC 7321, IC5249 and IC2233 were taken from \cite{banerjee2016mass}. The same for FGC1540 was obtianed from 
\cite{kurapati2018mass}. For UGC711, the same was modelled using using gipsy. In Table 1, we summarize the dark matter halo parameters of our sample superthins, with 
the stellar component modelled using the optical band. In Table 1, we summarize the same for the case in which the stellar component 
was modelled using the 3.6 $\mu$m photometry. We note that the dark matter parameters obtained for the above two cases are quite different. As discussed, the 3.6 $\mu$m band 
is a better representative of the stellar mass distribution. However, in order that our dynamical model is internally consistent, we use the the dark matter parameters 
from the mass models constructed using a given photometry as input parameters in the dynamical equations determining the structure and kinematics of the stellar disc in a 
given photometric band.

\section{Results \& Discussion}

\begin{figure*}
\resizebox{185mm}{195mm}{\includegraphics{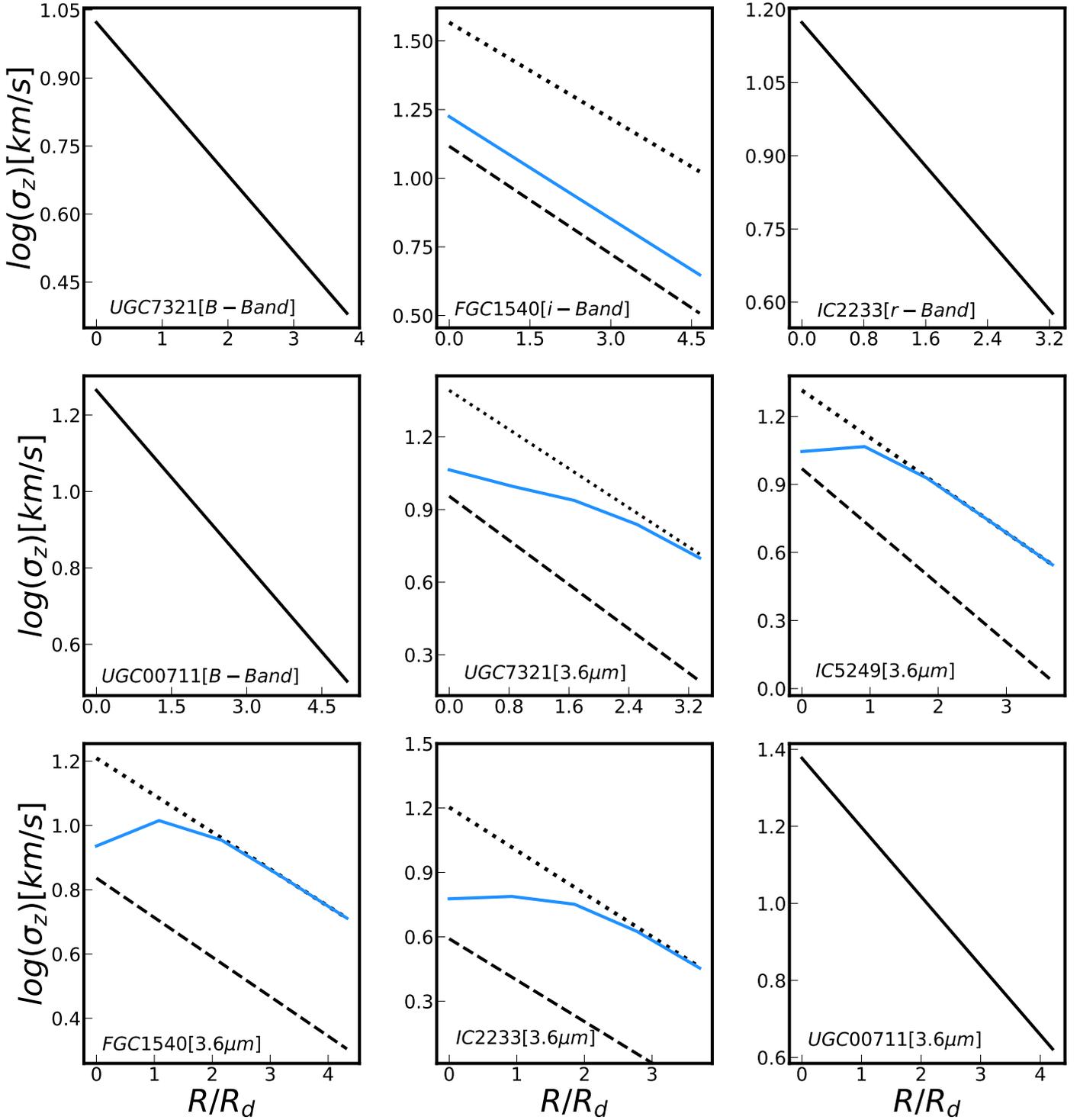}} 
\caption{We plot our model-predicted stellar vertical velocity dispersion 
in logarithmic scale for our sample superthins as a function of galacto-centric radius $R$ normalised by the exponential 
stellar disc scale length $R_{d}$. In case of 2-component stellar discs $R_{d}$ corresponds to $R_{d1}$ i.e., the scale length of 
the thick disc. In case of 1-component stellar disc, stellar vertical velocity dispersion is indicated by a single $“solid”$  
line. In case of 2-component discs, the $“dashed”$ line corresponds to the thin disc, the $“dotted”$ line to the thick disc and the $“solid”$ 
line the density-averaged value of the same.}
\end{figure*}

\begin{figure*}
\resizebox{185mm}{195mm}{\includegraphics{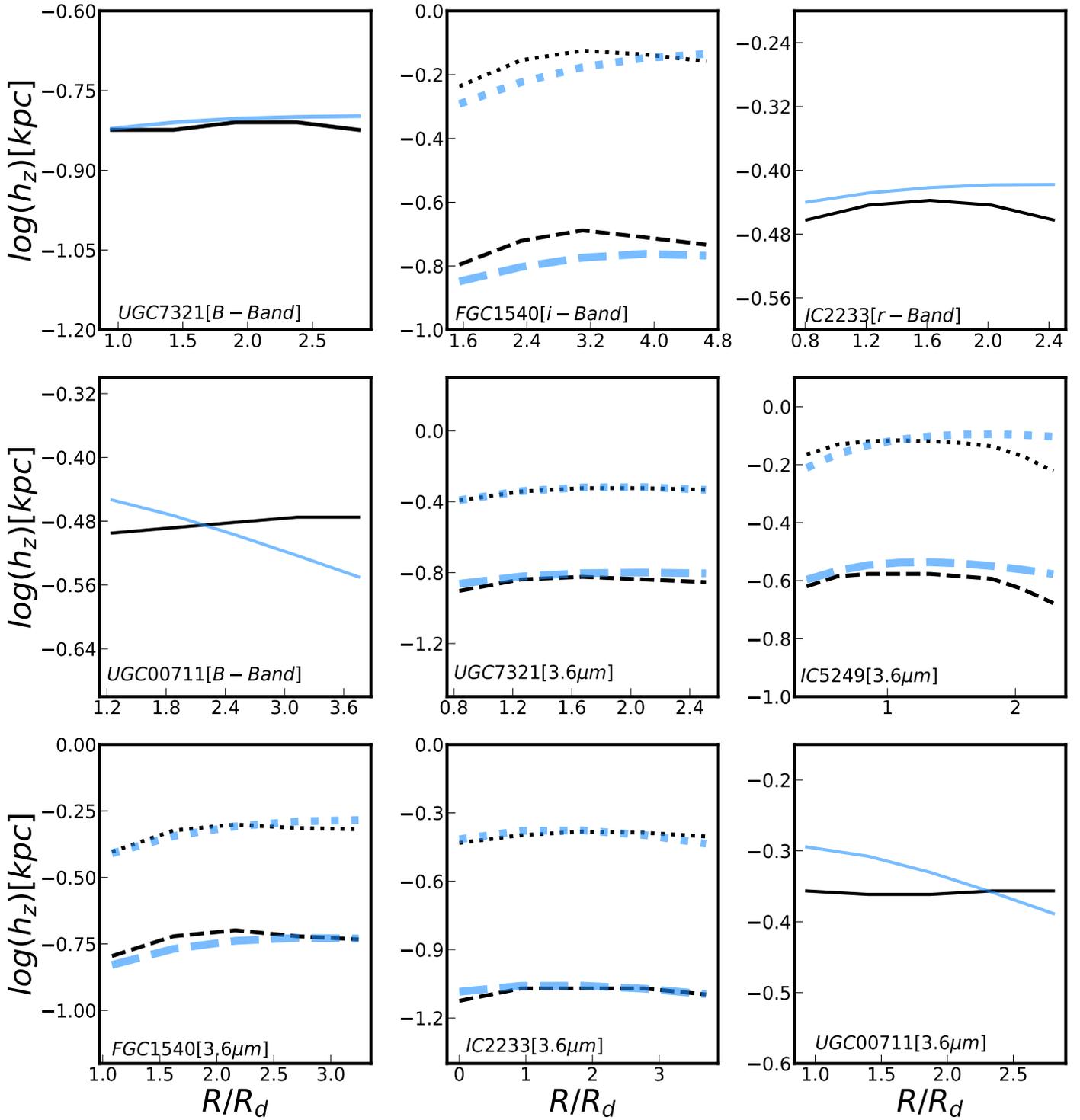}} 
\caption{We plot our model-predicted stellar scaleheight in logarithmic scale for our sample superthins as a function of galacto-centric radius $R$ normalised by the exponential stellar disc scale length $R_{d}$. In case of 2-component stellar discs $R_{d}$ corresponds to $R_{d1}$ i.e., the scale length of the thick disc.  In case of 1-component stellar disc, stellar vertical scale height is indicated
by a single $“solid”$  line. In case of 2-component discs, the $“dashed”$ line corresponds to the thin disc whereas the $“dotted”$ line to the thick disc.
The $“black”$ lines correspond to the multi-component model whereas the $“blue”$ lines to AGAMA.
}
\end{figure*}

\begin{figure*}
\resizebox{185mm}{195mm}{\includegraphics{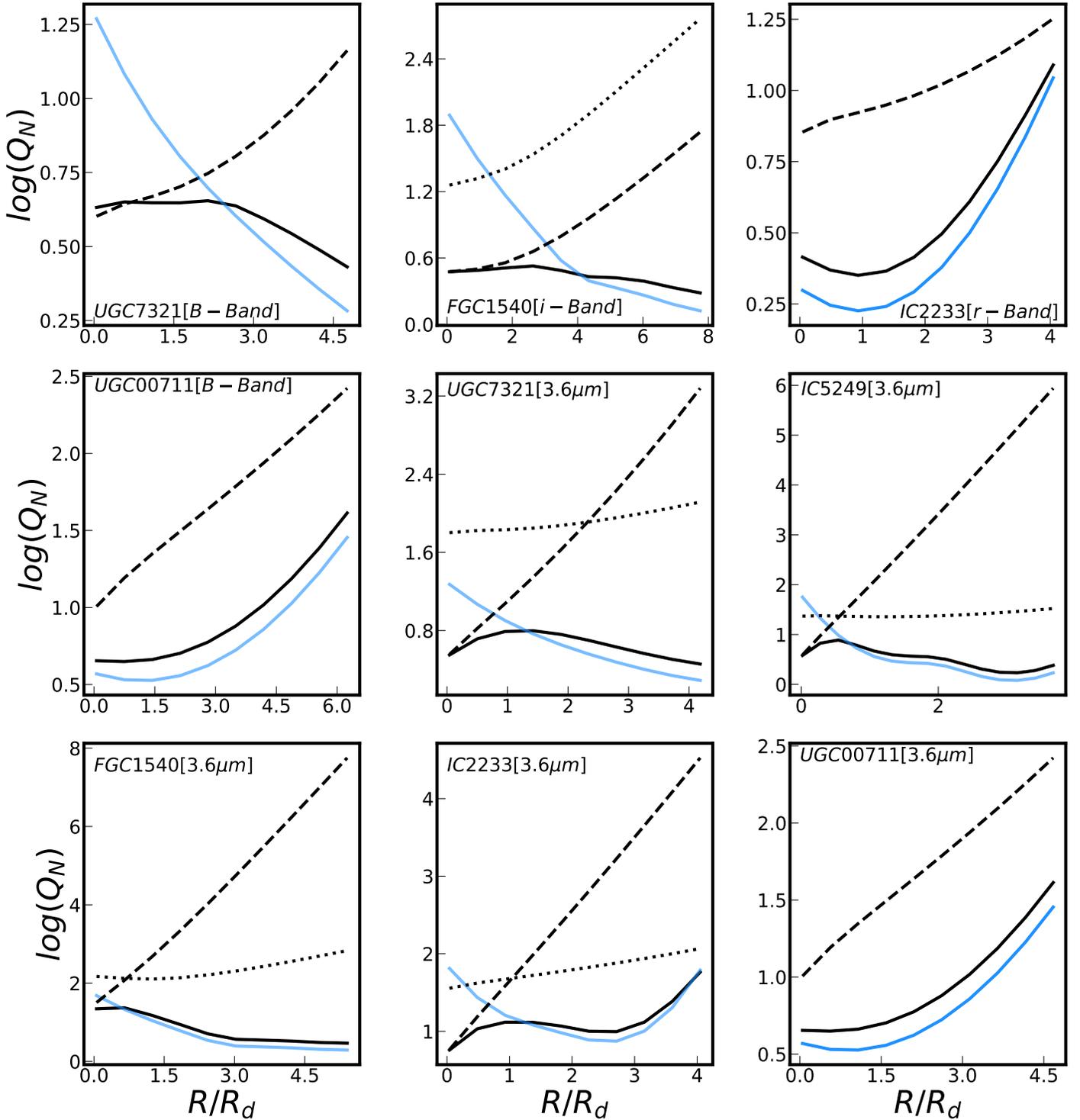}} 
\caption{
We plot our stabilty parameter $Q_{N}$ in logarithmic scale for our sample superthins as a function of galacto-centric radius $R$ normalised by the exponential stellar disc scale length $R_{d}$. In case of 2-component stellar discs $R_{d}$ corresponds to $R_{d1}$ i.e., the scale length of 
the thick disc. In case of 1-component stellar disc, the $“dashed”$ line corresponds to the Toomre Q of the stellar disc. In case of a 2-component stellar disc, the $“dashed”$ line corresponds to the Toomre Q of the thin disc whereas the $“dotted”$ line to the Toomre Q of the thick disc. In both cases, the $“blue”$ line corresponds to the Toomre Q of the gas disc whereas the $“solid”$ line the multi-component disc stability parameter $Q_{N}$.}
\end{figure*}

\begin{table*}
\begin{center}
\hfill{}
\caption{Best-fitting parameters of the multi-component models of our sample superthin galaxies.}

\begin{tabular}{|l|c|c|c|c|c|c|}
\hline
Parameters      & $\sigma^{\textcolor{red}{1}}_{0sI}$ &$\alpha^{\textcolor{red}{2}}_{sI}$ & $\sigma^{\textcolor{red}{3}}_{s0II}$& $\alpha^{\textcolor{red}{4}}_{sII}$ &$\sigma^{\textcolor{red}{5}}_{avg}$    \\
Galaxy          &km$s^{-1}$            &              &        km$s^{-1}$    &          &  km$s^{-1}$                  \\
\hline    
\hline
Vertical stellar dispersion:  Stellar disc modelled using optical photometry  .\\
\hline 
UGC 7321 & - & - & $10.23\pm 0.64$& $2.58 \pm 0.613$& \\
FGC 1540 &$36.91\pm 1.14$&$3.72 \pm 0.4$&$13.08\pm 1.17$& $3.32\pm 0.42$&16.78\\

IC  2233 &$14.9\pm 0.57$ & $2.36\pm 0.36$\\
UGC 00711&$18.4 \pm 0.87$& $3.21 \pm 0.40$\\
\hline
Vertical stellar dispersion: Stellar disc modelled 3.6 $\mu$m photometry.\\
\hline
UGC 7321 &$24.66 \pm 0.88$&$2.15 \pm 0.6$&$9.02 \pm 0.8$& $4.55 \pm 0.68$& 11.58\\ 
IC  5249 &$20.64\pm 0.63$&$2.155 \pm 0.217$&$9.32 \pm 0.39 $&$7.54 \pm 0.23$& 11.08\\
FGC 1540 &$16.20\pm0.87$&$3.77\pm 0.42$&$6.86 \pm 0.57$&$12.1\pm 0.59$&8.63\\
IC  2233 &$15.97 \pm 0.54$&$2.16 \pm 0.42$&$3.9 \pm 0.23$& $6.0 \pm 0.2$& 5.92\\
UGC 00711 &$23.82 \pm 1.45$&$2.42 \pm 0.28$\\ 
\hline
Vertical HI dispersion:  Stellar disc modelled using optical photometry  \\
\hline
Parameters&$\sigma^{\textcolor{red}{6}}_{0HI}$&$\alpha^{\textcolor{red}{7}}_{HI}$&$\beta^{\textcolor{red}{8}}_{HI}$\\ 
          &km$s^{-1}$                     \\
\hline

UGC 7321 & $11.06 \pm 0.88 $ & $0.18 \pm 0.07$  &$-0.047 \pm 0.02$\\ 
FGC 1540 & $29.01 \pm 1.16 $ & $4.27 \pm 0.425$ &            \\
IC  2233 & $12.52 \pm 0.515$ & $1.03 \pm 0.14$  &$-0.141 \pm 0.031$\\
UGC 00711& $23.10 \pm 1.11 $ & $1.03 \pm 0.145$ &$-0.156 \pm 0.05 $ \\ 
\hline
Vertical HI dispersion:  Stellar disc modelled 3.6 $\mu$m photometry.\\
\hline
UGC 7321 &$11.19 \pm 0.84$  & $-0.29 \pm0.14$&\\
IC  5249 &$12.4  \pm 0.53$  & $ -0.99 \pm 0.11$&$0.04 \pm 0.0011$\\
FGC 1540 &$17.75 \pm 0.81$  & $6.85 \pm 0.56$&  \\
IC  2233 &$12.0  \pm  0.56$ & $0.53 \pm 0.23$ &$-0.055 \pm 0.026$\\
UGC 00711&$22.03 \pm 1.07$  & $0.92 \pm 0.16$&$-0.1 \pm 0.054$\\
\hline
 \end{tabular}
\hfill{}
\label{table:results}
\end{center}
\begin{tablenotes}
\item (\textcolor{red}{1}): Central vertical stellar velocity dispersion in thick disc 
\item (\textcolor{red}{2}): Scale length of radial fall off of the thick disc stellar dispersion in units of $R_{d1}$ 
\item (\textcolor{red}{3}): Central vertical stellar velocity dispersion in the thin disc
\item (\textcolor{red}{4}): Scale length of radial fall off of the thin disc stellar dispersion in units of $R_{d2}$ 
\item (\textcolor{red}{5}): Average stellar dispersion
\item (\textcolor{red}{6}): Central vertical HI dispersion 
\item (\textcolor{red}{7}): steepness parameter-1 of HI dispersion profile
\item (\textcolor{red}{8}): steepness parameter-2 of HI dispersion profile
\end{tablenotes}
\end{table*}

\subsection{Stellar vertical velocity dispersion}

We present the results obtained from dynamical models of our sample of superthin galaxies as constrained by stellar photometry and HI 21cm radio-synthesis observations. In the figure 1, we present the vertical component of the stellar velocity dispersion in logarithmic scale $log \sigma_{z}$ as a function of $R/R_{d}$ obtained using the multi-component model of gravitationally-coupled stars and gas in the force field of the dark matter halo for our sample superthin galaxies. For each of our sample galaxies, we consider two different cases for the above dynamical model, depending on the choice of the photometric band to model the stellar component (i) Optical (ii) 3.6 $\mu$. Except for FGC1540, our sample superthins is composed of a thin exponential stellar disc only in the $optical$-band. FGC1540, on the other hand, is composed 
of a thin and a thick stellar disc in the optical band. We note that in the optical band, the central value of the vertical component of the stellar velocity dispersion $\sigma_{0s}($ varies between $10.2$ and $18.4$ kms$^{-1}$.

In the 3.6 $\mu$m band, the stellar component of all our sample galaxies, except for UGC711, is composed of a thin and a thick exponential stellar disc. UGC711 is composed of a thick disc only in the 3.6 $\mu$m band.  Interestingly, the central value of the vertical component of
the stellar velocity dispersion $\sigma_{0s}$ of the thin disc ranges between $3.9$ to $9.3$ kms$^{-1}$. The same for the thick disc varies 
between $15.9$ to $24.7$ kms$^{-1}$. The weighted average value for the thin and the thick  stellar disc for the same lies between $5.9$ to $11.6$ kms$^{-1}$. The optical disc constitutes young stellar population whereas the 3.6 $\mu$m  band traces the older stellar population. We note that the thin disc component in 3.6$\mu$m band  of UGC 7321, IC 5249, FGC 1540 and IC 2233 has a vertical stellar velocity dispersion, much lower than the corresponding value of the optical disc. We suspect that these galaxies might have undergone a recent star formation event, wherein the short lived  young stars have passed on to the red giant phase, emitting in near-infrared. Thus, the thin disc in 3.6 $\mu$m may represent this cold near-infrared component.

Previous studies on the vertical structure of disc galaxies mostly considered fixed values of $\alpha_{s}=2$, 
in order to comply with a stellar scale height constant with radius (See for example \cite{van1982surface}, \cite{van1988three}, \cite{van2011galaxy}, 
and also \cite{sharma2014kinematic}). Following \cite{narayan2002origin}), we consider $\alpha_{s}$ to be a free parameter. For the stellar discs in 
optical band, we find the value of $\alpha_{s}$ varies between 2.4 -3.7. For the 3.6 $\mu$m band, the thick disc $\alpha_{s}$ varies between 2.1 -3.7. 
These values are in compliance with \cite{narayan2002origin}, who found $\alpha_{s}$ to be lying between 2 and 3. However, for the 3.6 $\mu$m thin disc,
$\alpha_{s}$ lies between 4.5 -12.1.

In comparison to our sample superthins, the value of the central vertical stellar velocity dispersion 
in the Milky Way \citep{lewis1989kinematics} and Andromeda (M31) \citep{tamm2007visible} are $\sim$ 53 kms$^{-1}$, assuming that the dispersion 
falls off with a scale length of $2R_{d}$, and that the ratio of vertical-to-radial stellar velocity dispersion equals 0.5 in solar neighbourhood
\citep{binney2008princeton}. \cite{sharma2014kinematic} fitted  kinematic models to the Radial Velocity Experiment (RAVE)\citep{steinmetz2006radial} 
and  Geneva Copenhagen Survey (GCS) \citep{nordstrom2004geneva} data and found that the vertical velocity dispersion of the 
thin and thick disc of Milky Way stars are respectively given by  $\sigma_{0sII}=25.73^{+0.21}_{-0.21}$km/s and $\sigma_{0sI}=34.3^{+0.51}_{-0.57}$km/s, 
the exponential scale lengths being $1/(\alpha_{s2}R_{d2})=0.073^{+0.0037}_{-0.003}$ $kpc^{-1}$ and $1/(\alpha_{sI}R_{d1})=0.1328^{+0.005}_{-0.0051}$ $kpc^{-1}$ respectively. In fact, the central vertical dispersion values  of the older stellar population in two face-on spirals, NGC 628 and NGC 1566, are 60 kms$^{-1}$ and 80 kms$^{-1}$ respectively \cite{van1984vertical}, which are much higher than the range of 15 - 24 kms$^{-1}$ observed in the 3.6 $\mu$m thick disc of our sample superthins. Compared to  
our sample of superthin galaxies the sample of 30 low inclination galaxies studied as part of the DiskMass survey have a central vertical dispersion in range 25.9 kms$^{-1}$ to 108.5 kms$^{-1}$ \citep{martinsson2013diskmass}.

The multi-component model can simultaneously constrain the HI velocity dispersion of the galaxies as well. However, we had HI scale height
measurements only for two of our sample galaxies: UGC 7321 and IC 5249. For the rest of the sample superthins, we estimated the HI scale height
using a scaling relation between the FWHM and HI diameter (see \S 4). Except for FGC1540, the HI velocity dispersion values obtained from the two 
different models of the stellar disc are comparable within error bars. We find that the HI dispersion for UGC 7321 and IC 5249 are 11.2 km/s and 12.4 km/s 
respectively, which roughly complies with the observed mean dispersion values $11.7 \pm 2.3$ km/s of nearby galaxies from the high-resolution THINGS HI surveys \citep{mogotsi2016hi}. The central HI velocity dispersion of our superthin galaxy sample lies in range of 11 - 29 kms$^{-1}$. This roughly matches the findings of an earlier study of a sample of spiral galaxies in which the HI velocity dispersion was found to vary between 9 - 22 kms$^{-1}$ with a mean value of 11.7 kms$^{-1}$ \cite{mogotsi2016hi}. A similar value of HI velocity dispersion were observed by \cite{tamburro2009driving} $\sim 10 \pm 2 $km/s. 
We note that low values of HI dispersion are consistent with the cold neutral medium (CNM), having velocity dispersion in 
range 3.4 - 14.3 kms$^{-1}$\citep{ianjamasimanana2012shapes}. Similarly, HI dispersion values on the higher side might possibly reflect presence of warm neutral medium (WNM) characterised by values 10.4 - 43.2 kms$^{-1}$ \citep{ianjamasimanana2012shapes}.
Our calculations also indicate that the vertical velocity dispersion of the stellar disc is lower than that of the HI 
at some radii in some cases. It is, in general, not possible for the stars to have lower dispersion than the gas clouds in which 
they are formed as stars are collision less and therefore cannot dissipate energy through collisions. This possibly indicates that 
the thin disc stars were borne of very cold low dispersion molecular clouds, which could not be detected due to the very low metallicity of
superthin galaxies. In the self-consistent model of gravitationally-coupled stars and gas, the vertical velocity dispersion of any component is 
tightly constrained by its own observed scale height value. Our model results have been summarised in Table 2.

\subsection{Model stellar scaleheight : Multi-component model versus AGAMA}
In the figure 2, we check the consistency of the multi-component model with
the publicly-available stellar dynamical code AGAMA. Using the best-fitting value of the vertical stellar dispersion as obtained from the multi-component model as an input parameter in AGAMA, we find that the scale height predicted by AGAMA complies with that from the multi-component model. As discussed earlier, in contrast to the multi-component model, AGAMA determines the stellar radial velocity dispersion in addition to the vertical velocity dispersion, and thereby the ratio of the same 
as a function of galacto-centric radius. In the model constructed using optical band photometry for the stellar component, we find that the ratio of the vertical-to-planar velocity dispersion remains roughly constant at 0.5 within 3 Rd. On the other hand, in the model constructed using 3.6 micron photometry, our model indicates the following: 
For the thin disc, the ratio remains constant at 0.5 in UGC7321, IC5249 and UGC711; for FGC 1540, it varies between 0.5 and 0.3 whereas for IC2233 it remains constant at 0.3. 
For the thick disc, it remains constant at 0.5 in IC5249 and UGC711, at 0.3 in FGC1540 and varies between 0.4 and 0.3 in UGC7321 and IC2233. This is in line with the findings 
of \citet{Gerssen2012}, who showed that vertical-to-planar stellar velocity dispersion ratio decreases sharply from early-to-late-type galaxies. For an Sbc galaxy like the Milky Way, this value is $\sim$ 0.5. But for later-type galaxies like the superthins, it can be significantly closer to 0.3.
Finally, we use the values of the radial stellar dispersion from AGAMA for testing the dynamical stability of our sample superthin galaxies.

\subsection{ Disc dynamical stability}
In the figure 3, we plot the multi-component stability parameters $Q_N$ derived of \cite{romeo2013simple} as a function of $R/R_{d}$.
We further superpose the Toomre parameter for each component of the disc on the same plot. We note that except for the innermost galactocentric 
radii, $Q_{N}$ closely follows the Toomre parameter of the gas disc $Q_{g}$ for all our sample galaxies in general. 

\begin{figure}
\hspace*{-1.0cm}
\resizebox{90mm}{65mm}{\includegraphics{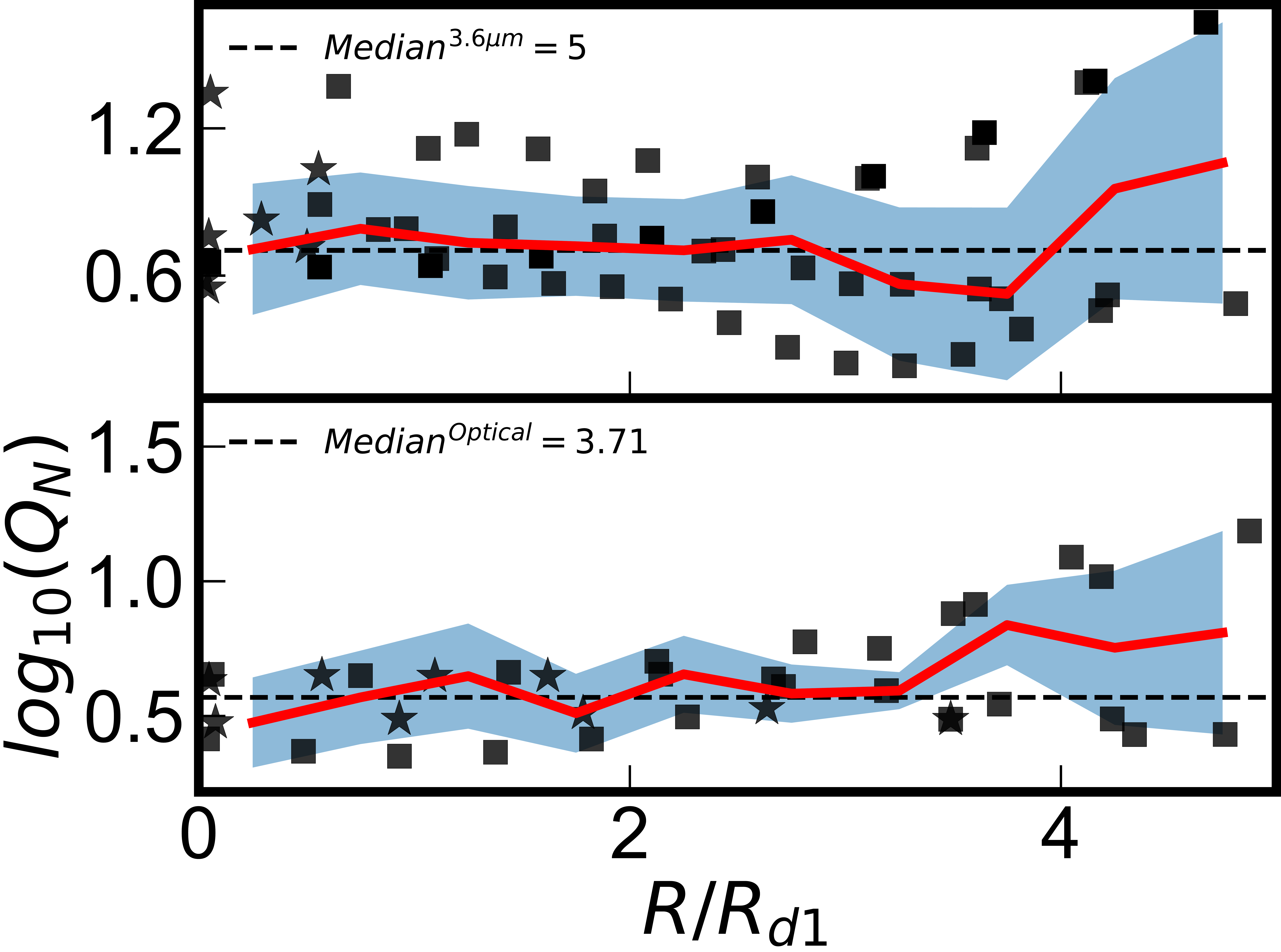}} 
\caption{The  figure  presents  the  multi-component disc stabilty parameter $Q_N$ of  the  our  sample  superthins as a function of galacto-centric
radius $R$ normalised by the exponential stellar disc scale length $R_{d}$. In case of 2-component stellar discs $R_{d}$ corresponds to $R_{d1}$ i.e., the scale length of 
the thick disc. Upper panel represents the case in which the stellar disc is modelled using 3.6 $\mu$ band whereas the lower panel the case corresponding to the optical band. 
$“Stars”$ indicate the $Q_{N}$ values driven by stellar disc and the $“squares”$ those driven by the gas disc. The solid red line indicates the local median of  the sample and 
the blue shaded region 1-sigma scatter about the median. The global median is indicated by the black dashed line.}
\end{figure}

In Figure 4, we present a composite plot of Log$Q_N$ versus $R/R_{d}$ for our sample superthins. The Upper and the Lower Panels correspond to the cases in which the 
stellar disc is modelled using 3.6 $\mu$m and optical photometry respectively. The local (solid line) and the global median (dotted) values of the sample are also indicated on the same plot. The blues shaded region represents the one sigma scatter of data points in each radial bin. The global median 
values for our superthin galaxies 5.0 $\pm$ 1.5 in the 3.6$\mu$m and 3.7 $\pm$ 1.5 in the optical band, 
these values are higher than the median $Q_{N}$ of 2.2 $\pm$ 0.6 found for a sample of nearby star-forming galaxies by 
\cite{romeo2017drives}, who also modelled their stellar discs using 3.6 $\mu$ photometry from the SINGS survey \citep{kennicutt2003sings}. 
This possibly indicates that the superthins are dynamically stabler than ordinary disc galaxies in general. We further note that the global median values 
for our superthin galaxies is higher than the  $Q_{critical}=2$ \citep{griv2012stability}, also higher than $Q_{critical}=2-3$ 
derived by \cite{elmegreen2011gravitational} taking into effect the destabilizing effect of gas dissipation. Our calculations show that in the 
optical band, the minimum $Q_N$ values of our sample superthins range between 1.9 and 4.5 with a median of 2.5. In the 3.6 $\mu$m band, 
the same varies between 1.7 and 5.7 with a median of 2.9. This complies with the median value 2.9 - 3.1 of the minimum $Q_N$ for a sample 
of low surface brightness galaxies as found in earlier studies \citep{garg2017origin}.

\begin{figure}
\hspace*{-1.0cm}
\resizebox{90mm}{65mm}{\includegraphics{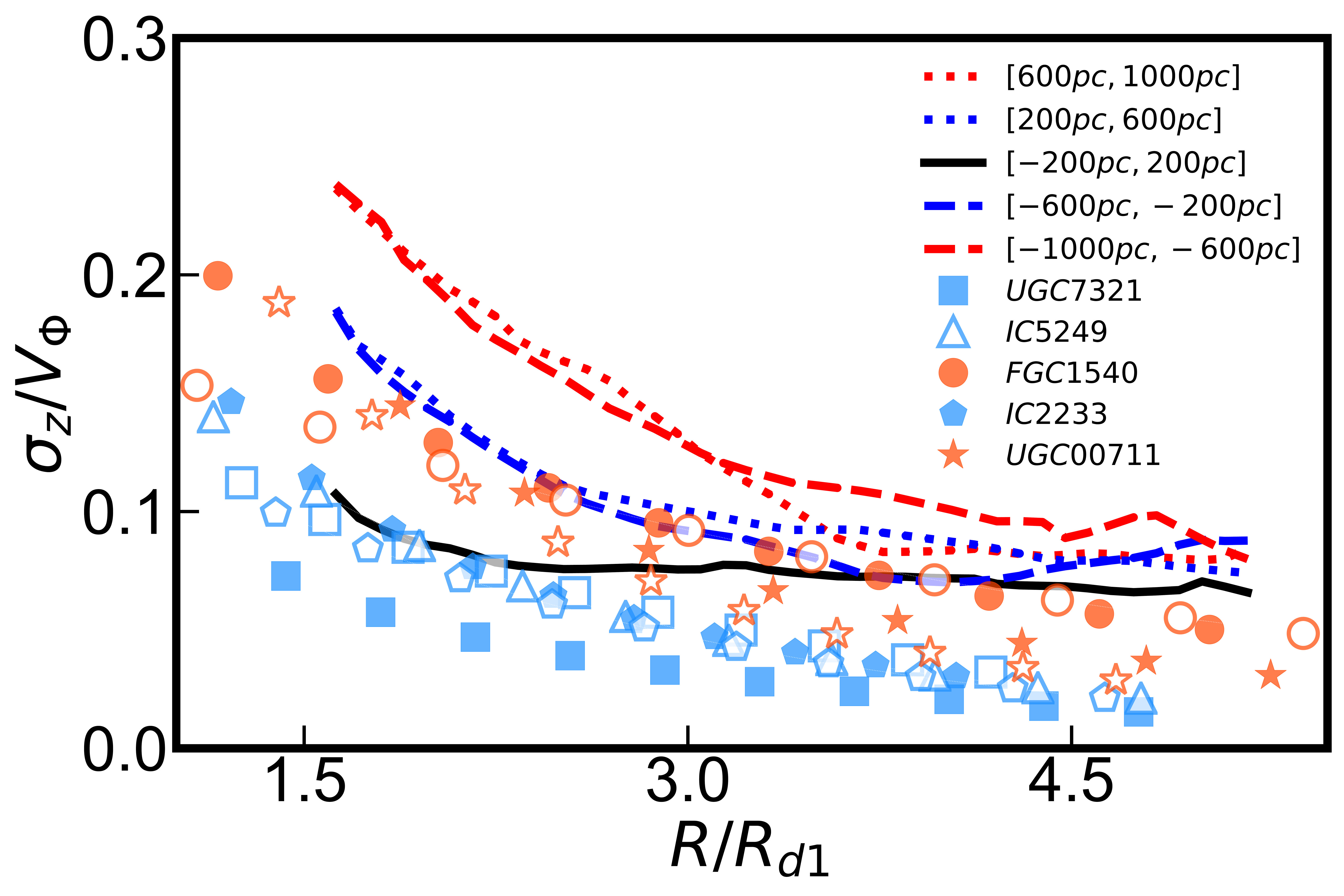}} 
\caption{We plot the ratio of the vertical stellar velocity dispersion to the asymptotic rotational velocity $\sigma_{z}/V_{\phi}$ versus
the galacto-centric radius $R$ normalised by $R_{d}$, the exponential stellar disc scale length for all our sample superthins. In case of 2-component 
stellar discs, $\sigma_{z}$  corresponds to the density-averaged vertical velocity dispersion of the thin and the thick discs, and $R_{d}$ corresponds to
the thick disc scale length. The filled markers represent the optical band stellar disc models while the unfilled markers for the same marker-style the 3.6$\mu$m models. 
The solid and dotted lines of different colours denote the $\sigma_{z}/V_{\phi}$ versus $R/R_{d}$ for the Milky Way stars corresponding to different vertical slices.}
\end{figure}

\subsection{How cold are superthin galaxies ?}
So far we have only compared the absolute values of the vertical velocity dispersion of superthin galaxies 
as predicted by our theoretical model with those of ordinary galaxies. We found that the superthin galaxies 
are cold systems as compared to the massive spiral galaxies like Milkyway in terms of absolute vertical velocity dispersion. 
In Figure 5, we plot the vertical velocity dispersion of the thin stellar disc, both in the optical and in the 3.6 $\mu$m band,
normalized by their rotational velocity as a function of $R=Rd_1$.We compare the same with the corresponding with the corresponding 
profile of the Milky Way stars lying within different vertical heights from the galactic mid-plane as taken from \cite{katz2018gaia}.
Except at the inner radii of UGC00711 and FGC1540, the value of $\sigma_{z}/V_{\phi}$ for all the superthin galaxies is lower than than 
the corresponding value for the Milky Way stars in -200 pc to 200pc vertical slice, indicating the superthin galaxies are indeed ultra-cold systems.

\textbf{Possible origin of the \emph{cold} stellar discs in superthins}\\
Bars, spiral arms, giant molecular clouds (GMCs) and satellite galaxies play an
important role in heating the galactic disc.
Spiral arms, on the other hand, heats the disc in the radial direction
\citep{aumer2016age}. GMCs heat the disc in both vertical and radial
direction \citep{jenkins1990spiral}.\cite{saha2014disc} showed that galaxies hosting
strong bars heat the disc very efficiently, leading to the formation of
thick discs. Using N-body simulations, \cite{grand2016vertical} have shown that the
time evolution of the bar strength correlates with the evolution of the global
vertical energy of the stellar particles. In contrast, superthin galaxies possibly have
weak bars which disfavour disc heating in the vertical direction and may thus
preserve the superthin vertical structure. \cite{aumer2016age} finds that massive
satellite galaxies and subhaloes also significantly heat the galactic disc in the
vertical direction.

\section{conclusions}
Superthin galaxies are a class of low surface brightness, bulgeless, disc galaxies, exhibiting sharp,
needle-like images in the optical, implying strikingly high values of planar-to-vertical axes ratios of
the stellar disc, which possibly indicates the presence of an ultra-cold stellar disc, the dynamical
stability of which continues to be a mystery. We construct dynamical models of a sample of
superthin galaxies using stellar photometry and HI 21cm radio-synthesis observations as
constraints and employing a Markov Chain Monte Carlo method, also checking the consistency of
our model results using AGAMA i.e. Action-based Galaxy Modelling Architecture \citep{vasiliev2018agama}.

\begin{itemize}
 \item We find that the central vertical velocity dispersion for the stellar disc in the optical band varies between $\sigma_{0s}$ $\sim$ $10.2 - 18.4$ $\rm{kms}^{-1}$  and falls off with an exponential scale length of $2.6$ to $3.2$ $R_{d}$ where $R_{d}$ is the exponential stellar disc scale length. Interestingly, in the 3.6 $\mu$m, the same, averaged over the two components of the stellar disc, varies between $5.9$ to $11.8$ $\rm{kms}^{-1}$, which is mainly representative of the denser, thinner and smaller of the two-disc components. However, the dispersion of the more massive disc component varies between $15.9 - 24.7$ with a scale length of $\sim$ 2.2 $R_{d}$. 
\emph{Our results are indicative of the presence of ultra-cold stellar discs in superthin galaxies.} 

\item We have constructed the stellar velocity ellipsoid 
($\sigma_{z}/\sigma_{R}$) by setting up an equilibrium distribution function for our sample superthins using AGAMA.We find that the value of 
$\sigma_{z}/\sigma_{R}$ lies in between 0.48 - 0.59 for optical models. $\sigma_{z}/\sigma_{R}$ for thick disc in 3.6 $\mu$m models varies between 0.4 - 0.55, the same for the thin disc lies between 0.28 - 0.51. 

\item Further, we find that the disc dynamical stability of super thin galaxies as indicated by $Q_{N}$ is mainly 
driven by the gas component. Finally, in the 3.6 $\mu$ band, the global median of the multi-component disc dynamical stability parameter $Q_N$ of 
our sample superthins is found to be 5 $\pm$ 1.5, which higher than the global median value of 2.2 $\pm$ 0.6 \citep{romeo2017drives} for a sample of spiral galaxies.

\end{itemize}

\section{Data Availibility}
The data underlying this article are available in the article itself.
\section{Acknowledgement}
AB would like to thank Prof. Francoise Combes for useful discussion, and Prof. Chanda Jog and Dr. Eugene Vasilev for useful 
comments and suggestions. AB would also acknowledge the research grant from DST-INSPIRE Faculty Fellowship (DST/INSPIRE/04/2014/015709) 
for partially supporting this work.

\subsection*{Softwares/Packages}
We have used publically available R and python packages in this work. We have used FME \citep{soetaert2010inverse} for MCMC modelling, for analysis 
of results we relied on packages ggmcmc\citep{fernandez2016ggmcmc}, BayesianTools \citep{hartig2017bayesiantools}, and for purpose of plotting we 
have used ggplot2\citep{wickham2011ggplot2}, Matplotlib\citep{hunter2007matplotlib},tonic \citep{vaughan2016false}. 

\small{\bibliographystyle{mnras}}
\bibliography{superthin_check,superthin-20JUN20}

\section{APPENDIX- A}
We will present the results pertaining to each galaxy constituting our sample superthin galaxies.
\begin{figure*}
\begin{center}
\begin{tabular}{cc} 
\resizebox{90mm}{85mm}{\includegraphics{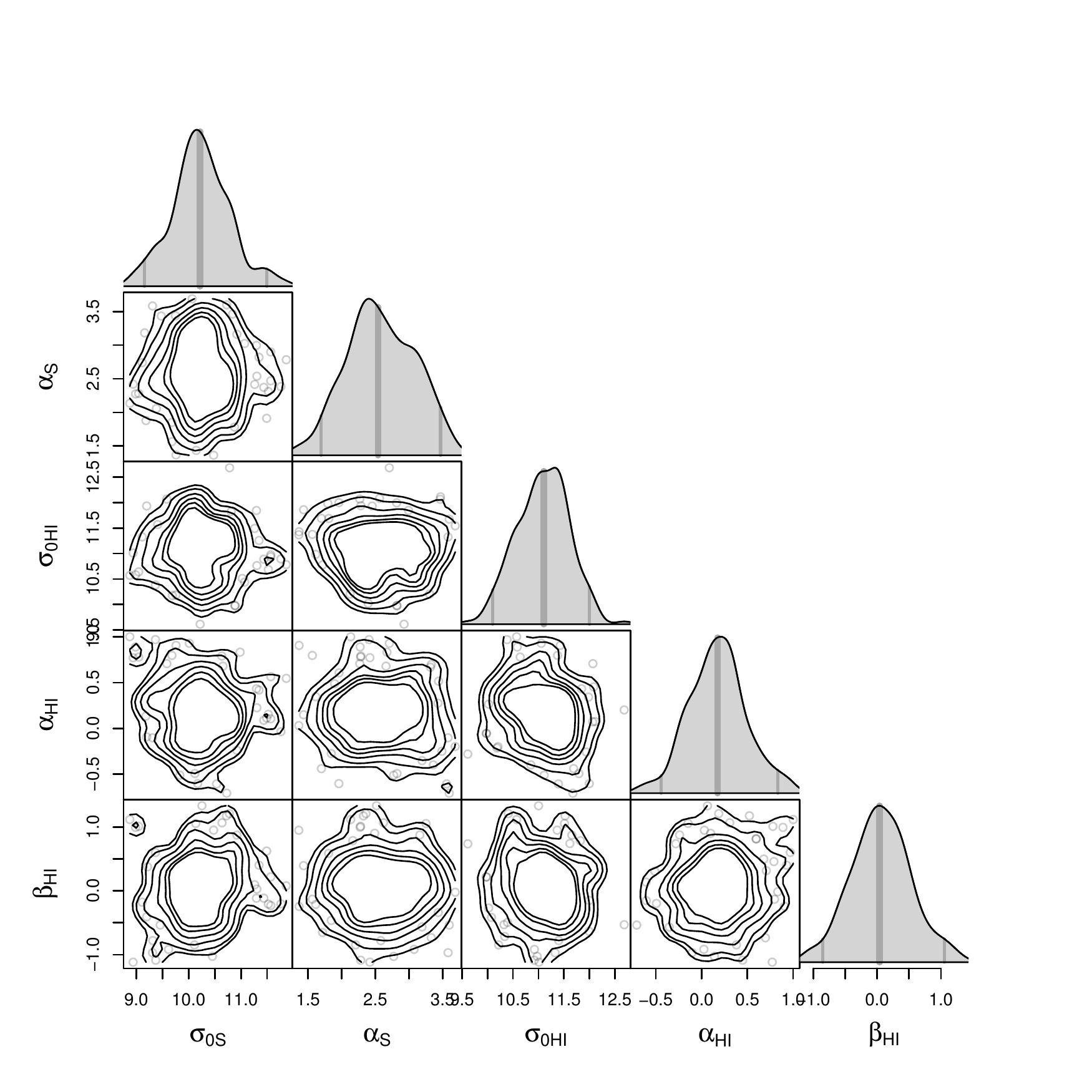}} &
\resizebox{90mm}{85mm}{\includegraphics{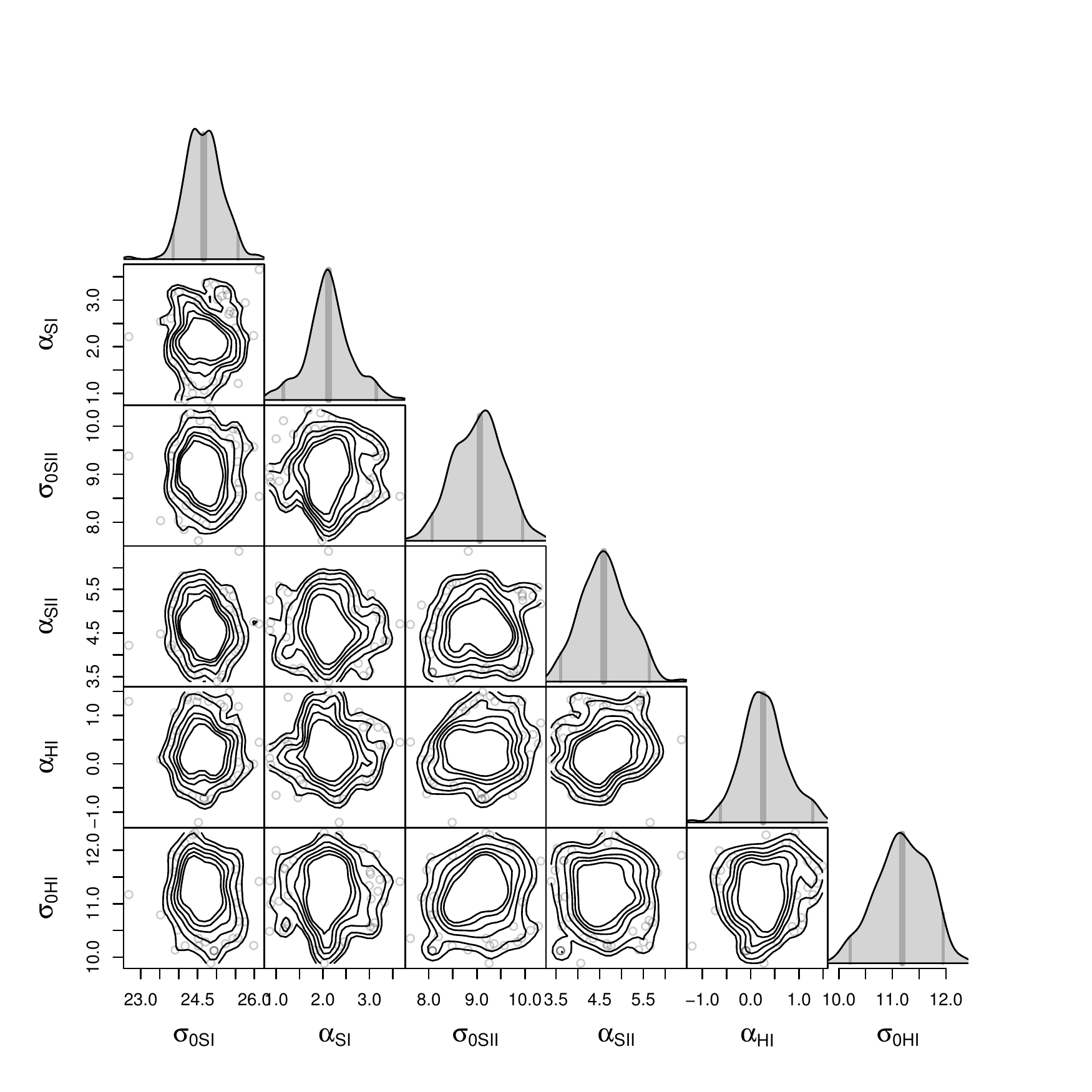}}\\
\end{tabular}
\end{center}
\caption{Posterior probability distribution and covariance plots of the parameters of the multi-component model of the galactic disc of UGC7321 with the stellar component 
modelled by $B$-band [Left Panel] and 3.6$\mu$m photometry [Right Panel]}
\end{figure*}
\begin{figure}
\begin{tabular}{c}
\resizebox{90mm}{85mm}{\includegraphics{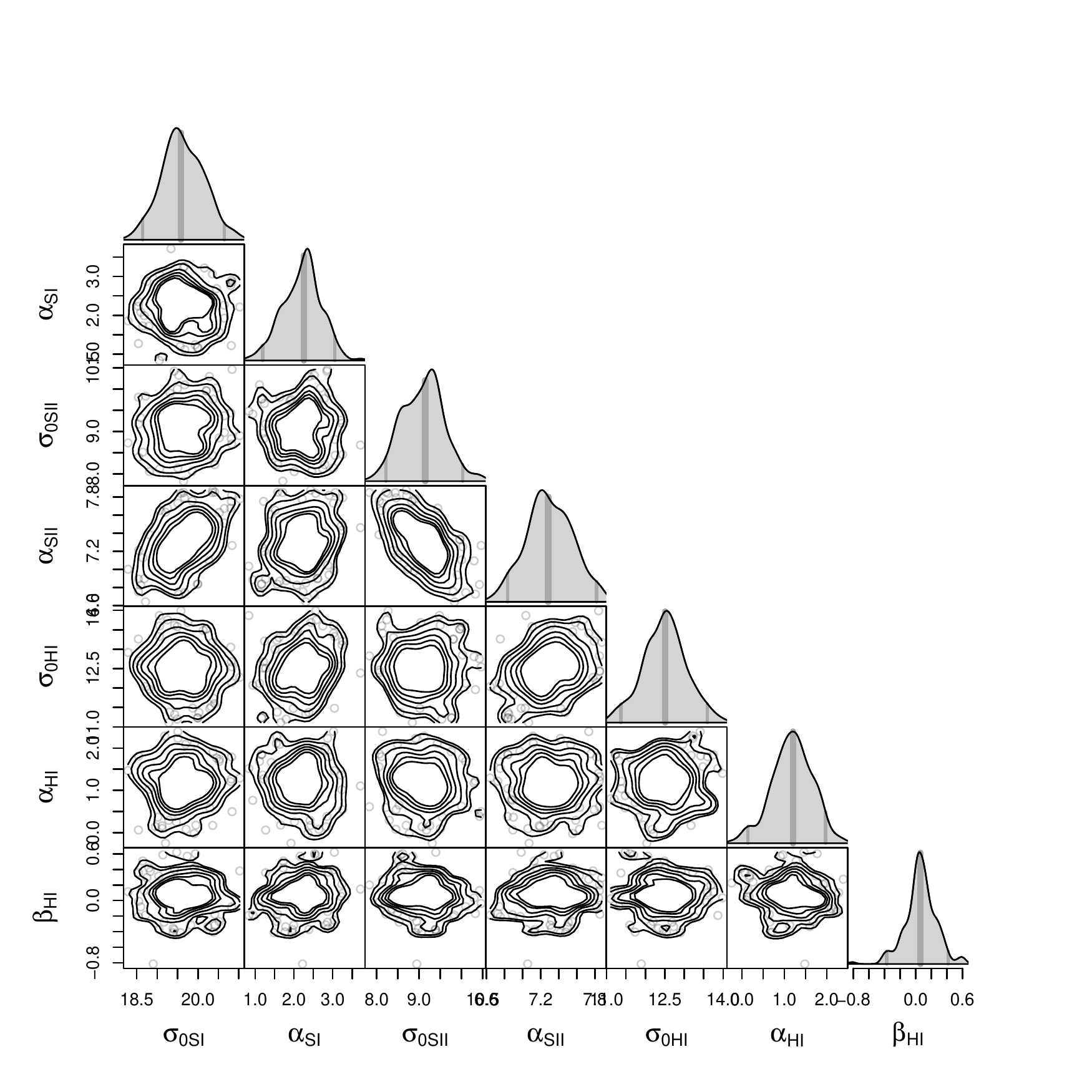}} 
\end{tabular}
\caption{Posterior probability distribution and covariance plots of the parameters of the multi-component model of the galactic disc of IC5249 with the stellar component modelled by 3.6$\mu$m photometry}
\end{figure}
\begin{figure*}
\begin{tabular}{cc}
\resizebox{90mm}{85mm}{\includegraphics{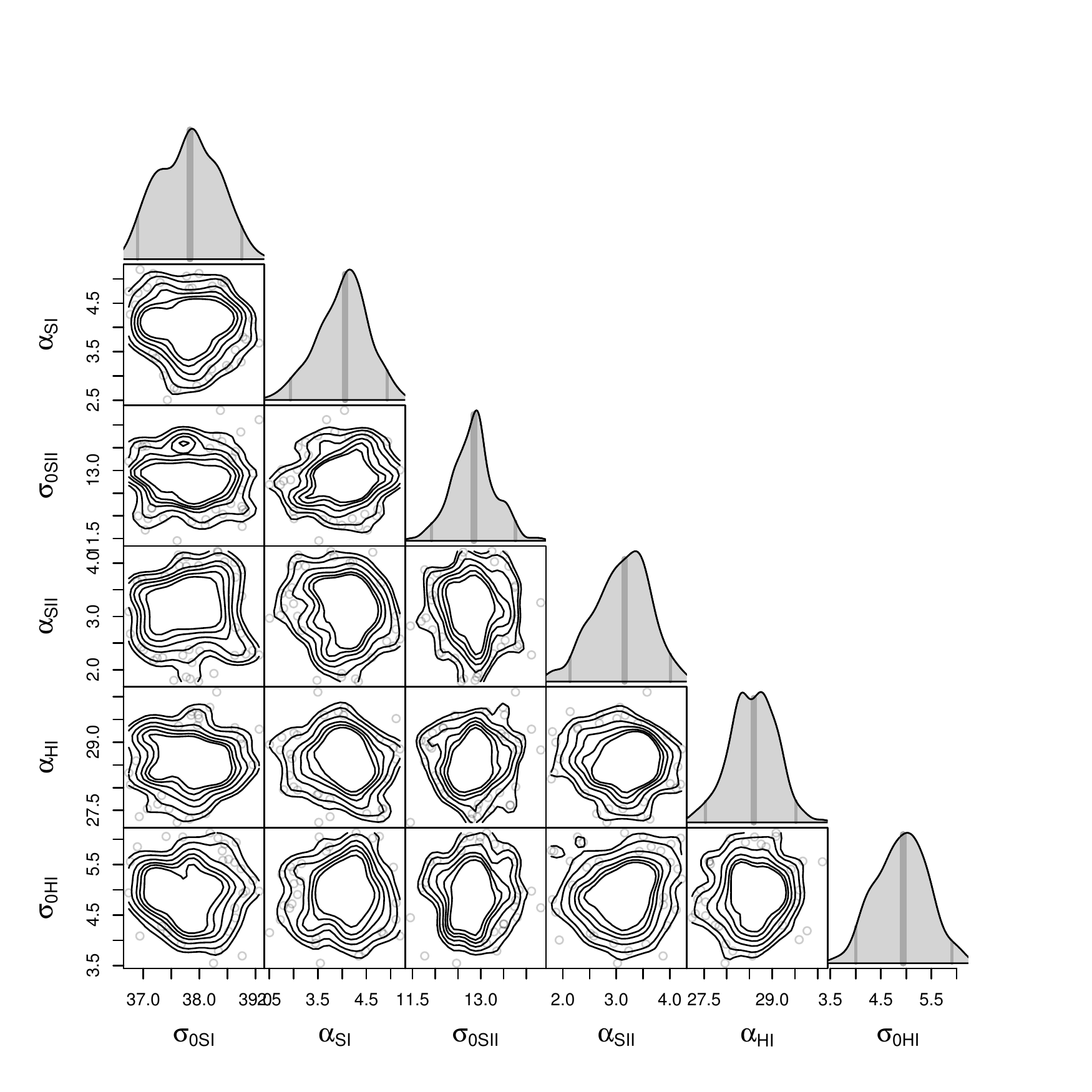}}&
\resizebox{90mm}{85mm}{\includegraphics{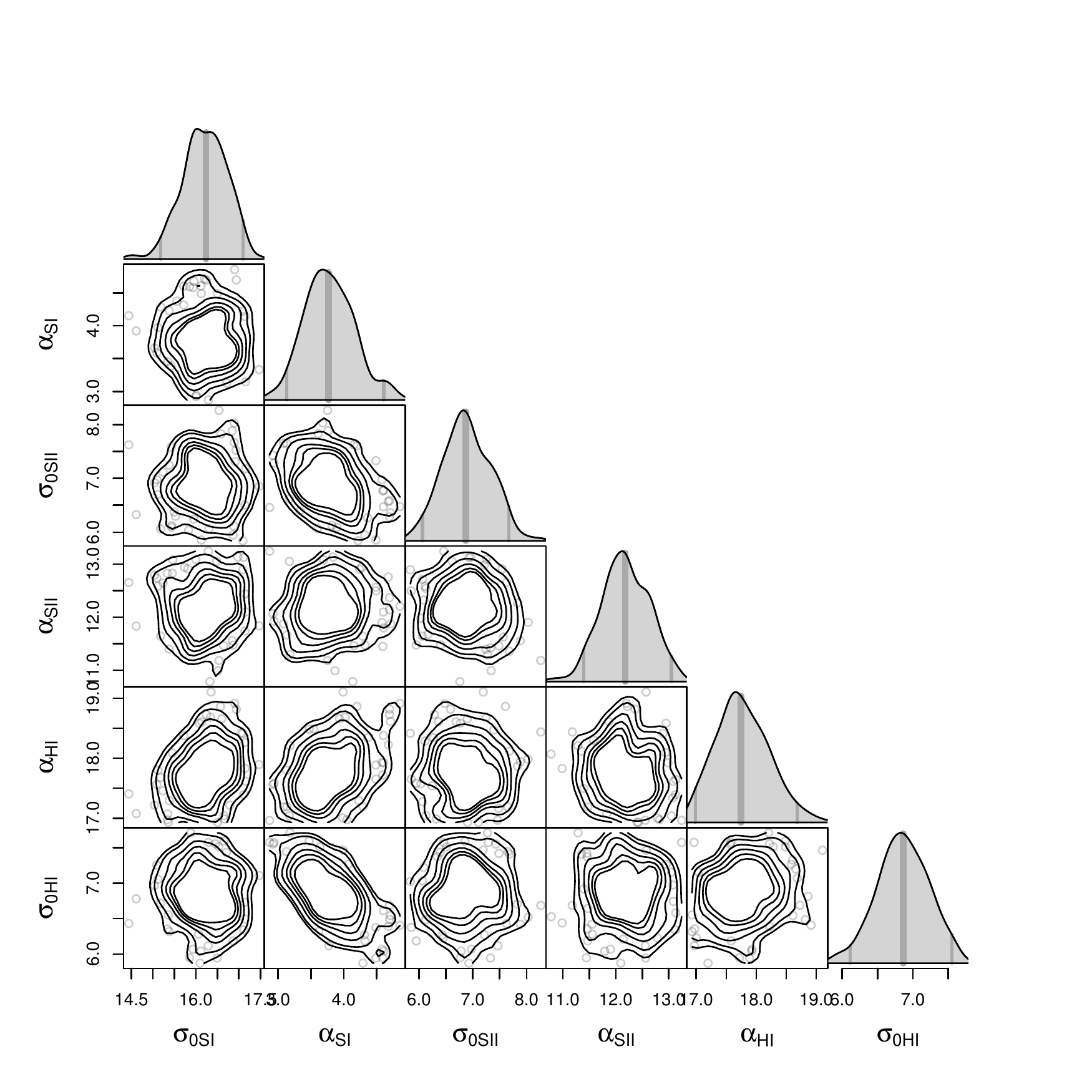}} 
\end{tabular}
\caption{Posterior probability distribution and covariance plots of the parameters of the multi-component model of the galactic disc of FGC1540 with the stellar 
component modelled by $i$-band [Left Panel] and 3.6$\mu$m photometry [Right Panel]}
\end{figure*}
\begin{figure*}
\begin{center}
\begin{tabular}{cc}
\resizebox{90mm}{85mm}{\includegraphics{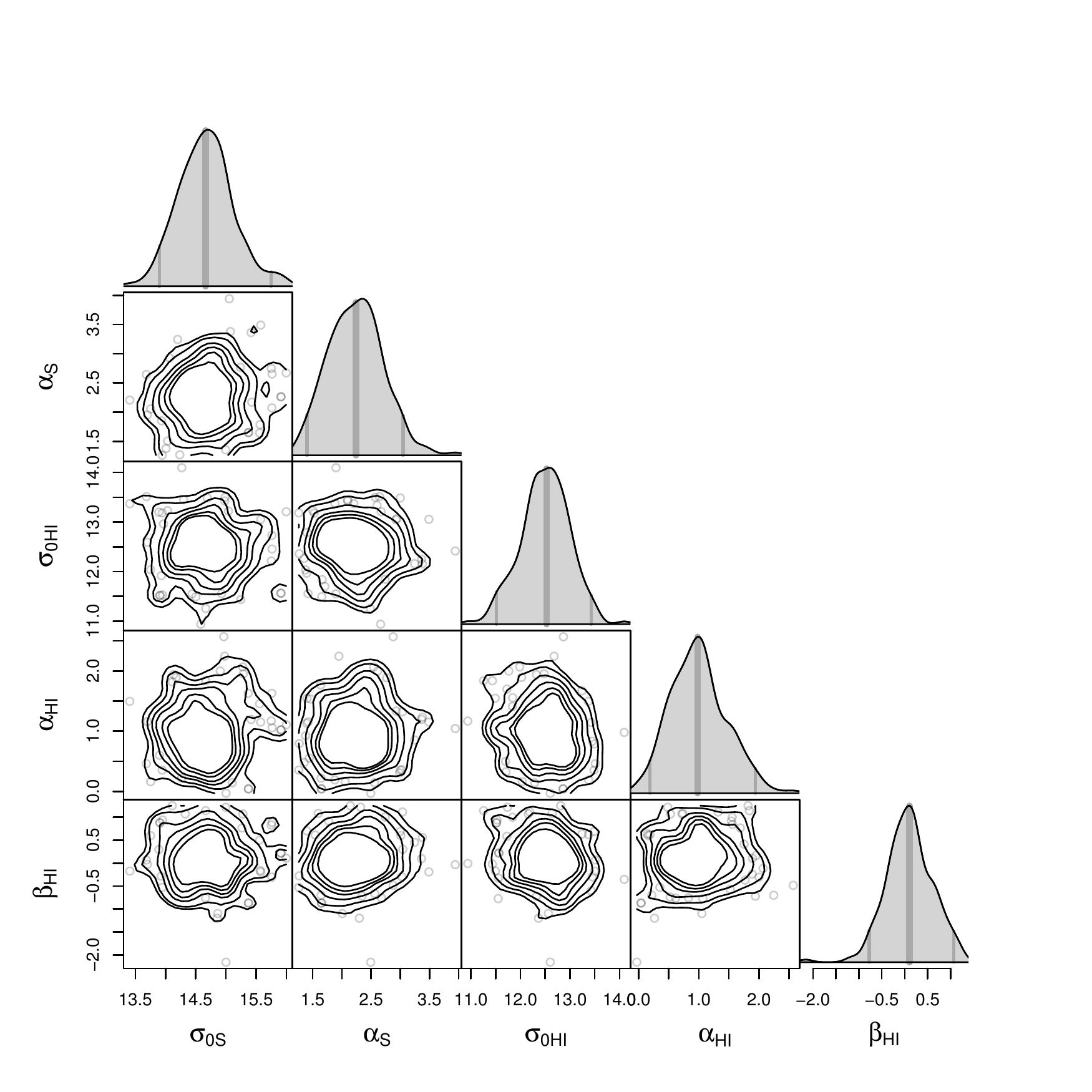}}&
\resizebox{90mm}{85mm}{\includegraphics{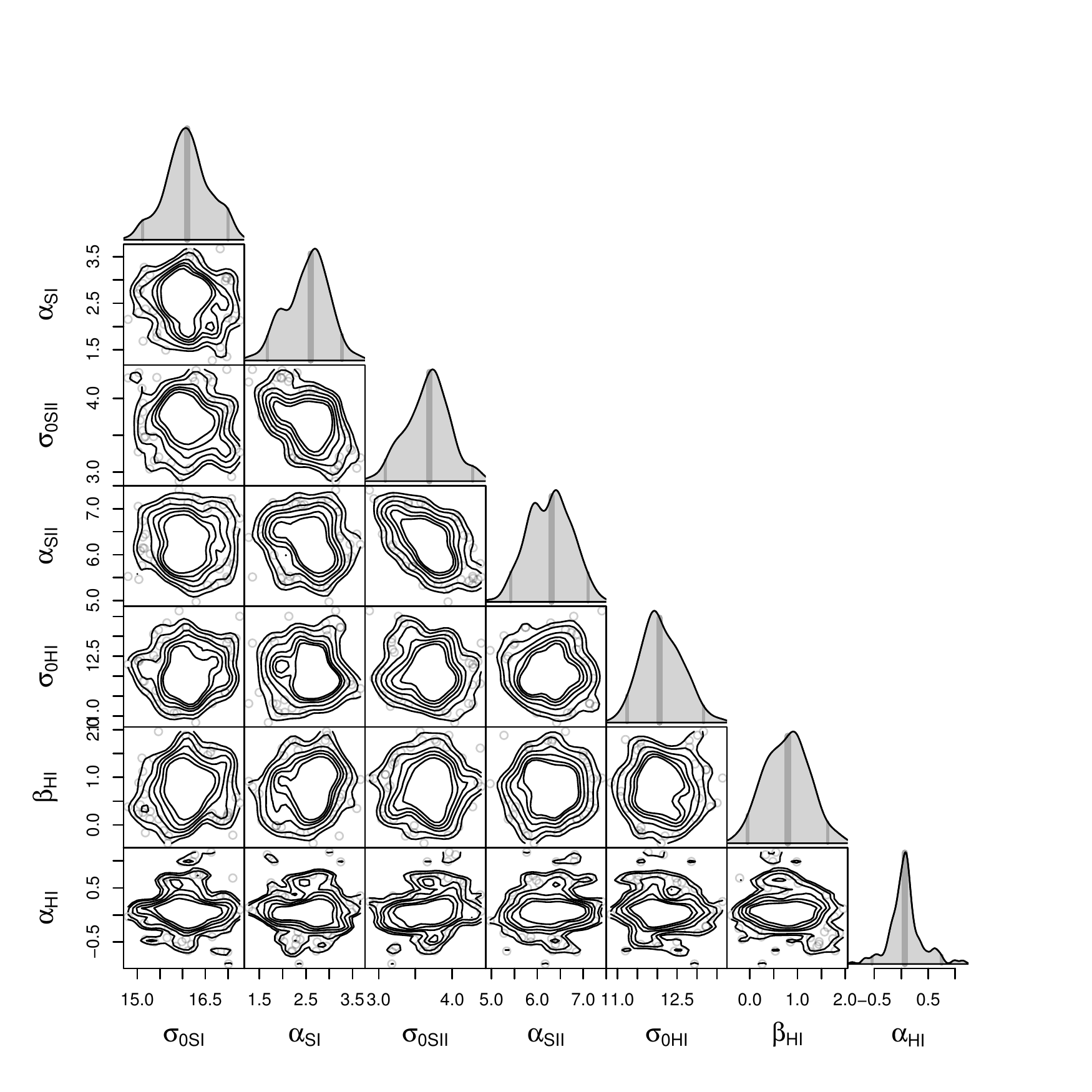}}
\end{tabular}
\end{center}
\caption{Posterior probability distribution and covariance plots of the parameters of the multi-component model of the galactic disc of IC2233 with the stellar component modelled by $r$-band [Left Panel] and 3.6$\mu$m photometry [Right Panel]}
\end{figure*}
\begin{figure*}
\begin{tabular}{cc}
\resizebox{90mm}{85mm}{\includegraphics{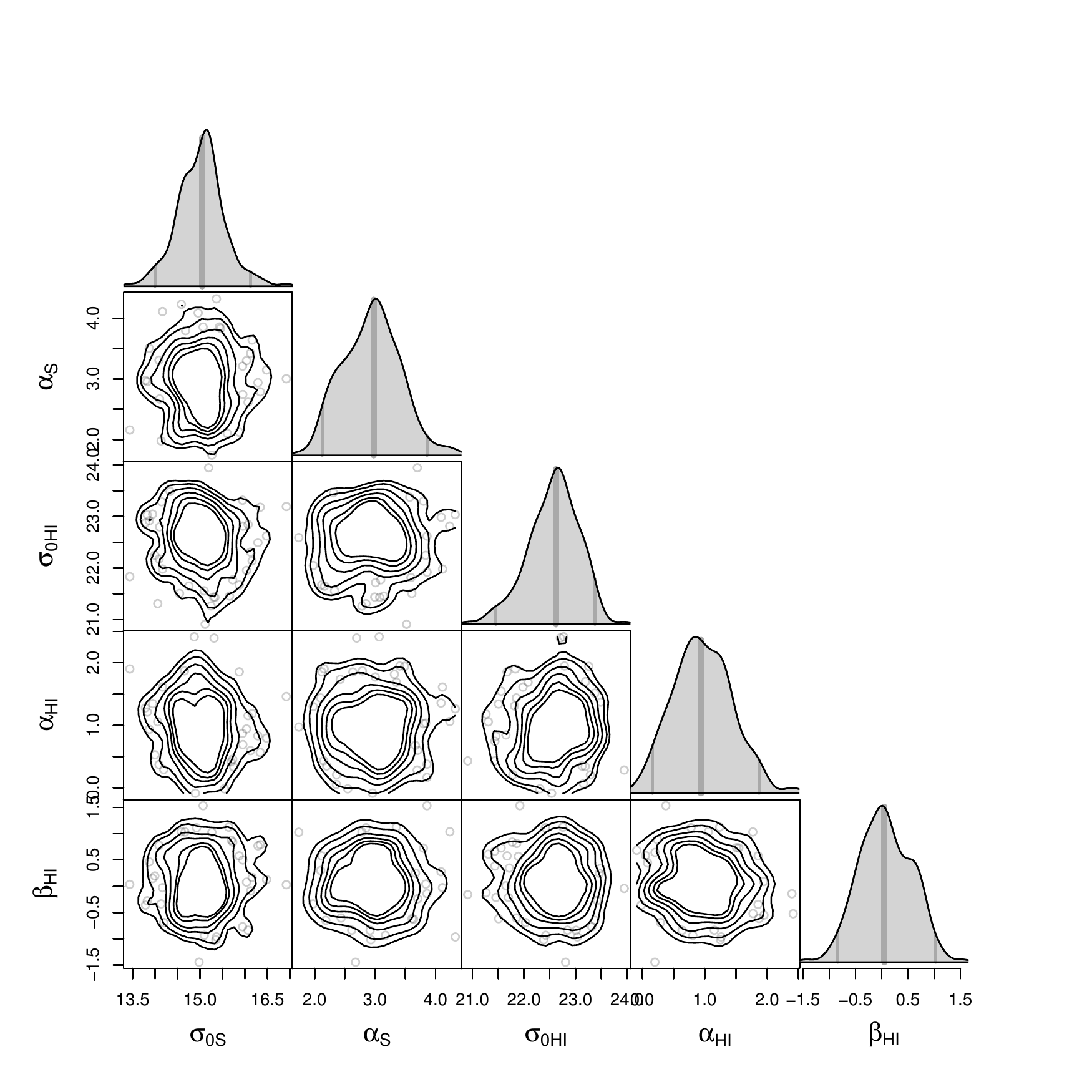}} &
\resizebox{90mm}{85mm}{\includegraphics{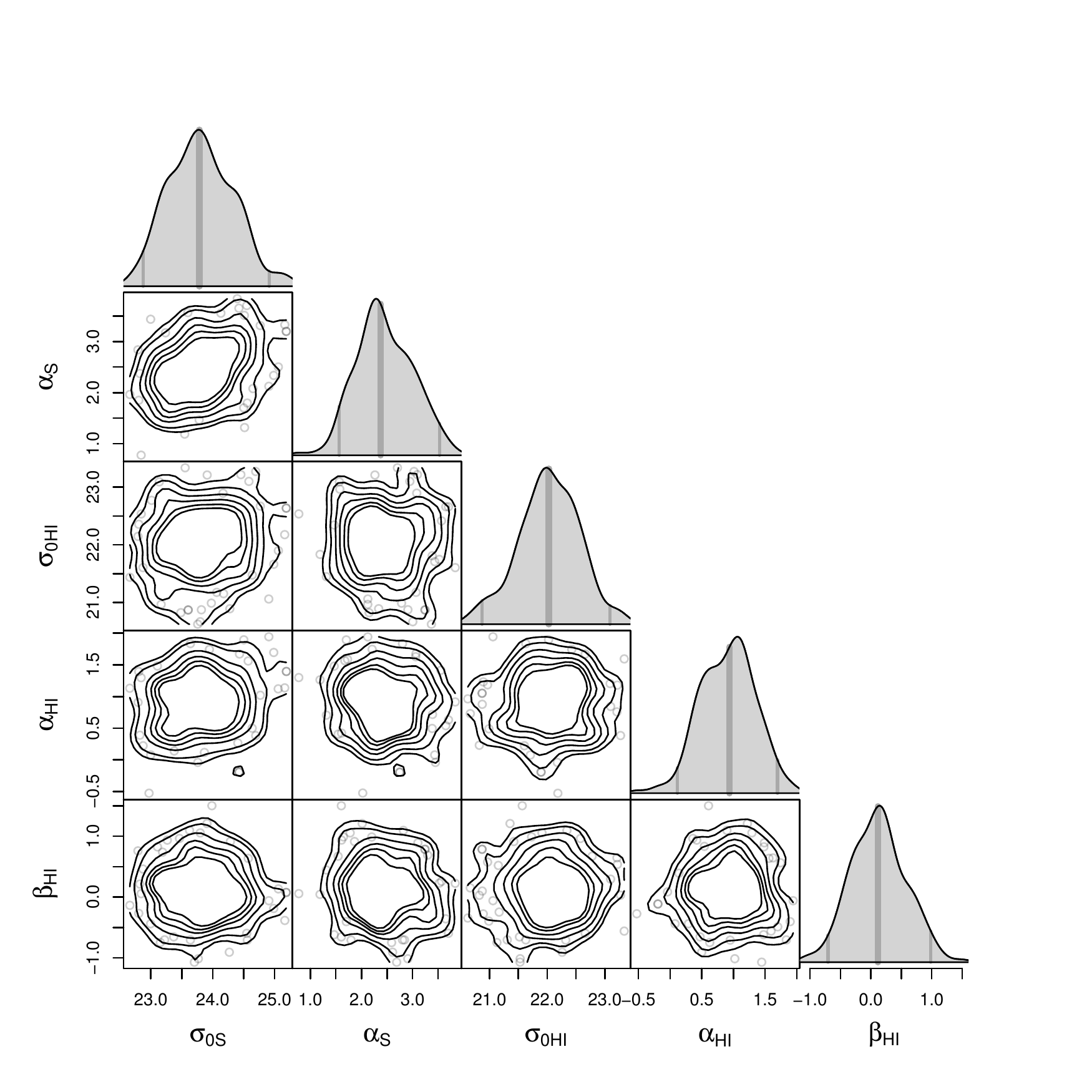}}\\
\end{tabular}
\caption{Posterior probability distribution and covariance plots of the parameters of the multi-component model of the galactic disc of UGC00711 with the stellar component modelled by $B$-band [Left Panel] and 3.6$\mu$m photometry [Right Panel]}
\end{figure*}

\subsection{UGC7321}

We describe the results obtained from dynamical modeling of UGC 7321 as constrained by stellar photometry in the $B$-band in addition to 
HI 21cm radio-synthesis observations. The central stellar dispersion $\sigma_{0s}$ = (10.2 $\pm$ 0.6) kms$^{-1}$, which falls off exponentially with a scalelength  
of (2.6$\pm$0.6)$R_{D}$. In comparison, the central value of the vertical velocity dispersion of the stellar disc in the Milky Way \cite{lewis1989kinematics} and the Andromeda or 
M31 \citep{tamm2007visible} is about $\sim$ 53 kms$^{-1}$. This is assuming that the stellar radial velocity dispersion falls off exponentially with a scale length of 2 Rd as 
is observed in the Galaxy, and also the ratio of the vertical to the radial stellar velocity dispersion is 0.5 at all radii, equal to its observed value in 
the solar neighbourhood \citep{binney2008princeton}. This confirms that UGC7321 has an ultra-cold stellar disc with unusually low values of the vertical velocity dispersion of stars. The central value of HI dispersion 
11.1 $\pm$0.9 kms$^{-1}$ with $\alpha_{HI}$=0.2$\pm$0.1 and $\beta_{HI}$=-0.04$\pm$0.02, thus indicating that the HI dispersion remains almost constant with 
$R$. We check the consistency of the multi-component model with the publicly-available stellar dynamical code AGAMA. Using the best-fitting value of the vertical stellar dispersion 
as obtained from the multi-component model as an input to AGAMA, we find that the scaleheight predicted by AGAMA complies with that from the multi-component 
model. Finally, we calculate multi-component disc dynamical stability parameter $Q_{RW}$ as function of $R$. We find that the minimum value of $Q_{RW}$ is 2.7 
at about 5$R_{D}$, thus confirming that UGC 7321 is stable against the growth of axisymmetric perturbations inspite of having an ultra-cold stellar disc. 
The central value of dispersion for the thin disc is 9.02$\pm$0.8 kms$^{-1}$, and it falls off exponentially with scale length (4.6$\pm$0.7)$R_{d2}$,
where $R_{d2}$ is the scale length of the thin disc. Interestingly,  the vertical velocity dispersion profile of the thin disc of the 3.6 $\mu$m stellar component almost matches 
the profile from the $B$-band component within error bars. The central value for the thick disc is 24.7$\pm$0.9 kms$^{-1}$, falling off exponentially with disc scale length 
(2.2$\pm$0.6)$R_{d1}$, $R_{d1}$ being the  scale length of the thick disc.  At small $R$, the density averaged vertical velocity dispersion is reflective of the cold, dense and 
compact thin disc. The central value of HI dispersion 11.2 $\pm$0.8 kms$^{-1}$ with $\alpha_{HI}$= -0.3$\pm$0.8 and $\beta_{HI}$=-0.04$\pm$0.02, again indicating that 
the HI dispersion remains almost constant with $R$. We note that the vertical velocity dispersion profile of HI as obtained from two models using different tracers for 
the stellar disc are comparable. In Figure 4, we present the posterior probability distribution of the parameters of the multi-component model of the 
galactic disc of UGC7321 with the stellar component modelled by $B$-band [Left Panel] and 3.6$\mu$m photometry [Right Panel].

\subsection{IC5249}
Our calculations show that the central value of the stellar vertical velocity dispersion of the thick disc is 20.6$\pm$0.6 kms$^{-1}$, which falls off exponentially 
with scale length (2.2 $\pm$ 0.2) $R_{d1}$. The same for the thin disc is found to be 9.3 $\pm$0.4 kms$^{-1}$ with an exponential fall-off scale length of (7.5 $\pm$0.2) $R_{d2}$. 
At $R > R_{d1}$, the density averaged vertical velocity dispersion converges to the value of the vertical velocity dispersion profile of the thick disc, which is hot, diffuse and 
extended. The central value of HI dispersion is 12.4 $\pm$0.5 kms$^{-1}$ with $\alpha_{HI}$= - 0.9 $\pm$0.1 and $\beta_{HI}$=-0.04$\pm$0.01, thus indicating that the HI dispersion 
remains almost constant with $R$. Although the results from the two models for the thin disc are fairly comparable, the same does not seem to hold
true for the thick disc, possibly because of its large scale height value. We find that the minimum value of $Q_{N}$ 
is 1.7 at about 3 $R_{d1}$, indicating that IC5249 may be on the borderline as far as dynamical stability of the stellar disc is concerned. 
In Figure 5, we present the posterior probability distribution of the parameters of the multi-component model of the galactic disc of IC5249 
with the stellar component modelled by the 3.6$\mu$m photometry 

\subsection{FGC1540}
Unlike our other sample galaxies, FGC1540 has two stellar discs in the optical. The central value of the vertical velocity dispersion of the thick disc is 
36.9 $\pm$1.1 kms$^{-1}$ and  falls off with an exponential scale length of 3.7 $\pm$ 0.4 $R_{d1}$. The central dispersion for the thin disc is 13.1 $\pm$1.2 kms$^{-1}$, the 
scale length of exponential fall-off being 3.3 $\pm$ 0.4 $R_{d2}$, where $R_{d2}$. The density averaged dispersion is reflective of the value of the thin disc component 
at all $R$. Due to the unavailability of the HI scale height data, we use a constant HI scale height of about 400 $kpc$ at all radius (Kurapati, private communication)
We find that the results for the thin disc are fairly comparable. However, as before, the results for the thick disc, which seem to be hotter than most of our sample galaxies, 
do not seem to be consistent with each other. We calculate $Q_N$ as a function of 
$R$, indicating a minimum value of 1.9.
The central vertical velocity dispersion of the thick disc is 16.2$\pm$0.9 kms$^{-1}$ and falls off exponentially with scale lenght 3.8 $\pm$0.4  $R_{d1}$.
The same for the thin disc is 6.9 $\pm$ 0.6 kms$^{-1}$, with a scale length of 12.1$\pm$0.2 $R_{d2}$. 
The density averaged dispersion converges with the thick disc value at large $R$. 
We note that the vertical velocity dispersion profiles of the thin disc in the optical fairly band matches with the thick disc in the 3.6 $\mu$m band. 
However, as discussed earlier, disparity in their central surface density values possibly rules out the fact that they trace the same disc. 
The minimum value of $Q_{N}$ is 2.9, indicating the disc can resist the growth of axis-symmetric instabilities, 
In Figure 6, we present the posterior probability distribution of the parameters of the multi-component model of the galactic disc of FGC1540 in 
$i$-band [Left Panel] and 3.6$\mu$m photometry [Right Panel].

\subsection{IC2233}
The central velocity dispersion of IC2233 in r-band is 14.9$\pm$0.6 kms$^{-1}$ and a scalelength of exponential fall-off equal to 2.4$\pm$0.4. We have calculated the $Q_{N}$ as 
function of $R$ which has a minimum value of $Q_{N}$ $\sim$ 2.2, confirming the stability of the disc against growth of axi-symmetric instabilites.  
The central vertical velocity dispersion of the thick disc in 3.6$\mu$m is 15.9 $\pm$0.5 kms$^{-1}$ and falls off exponentially with a scale length of (2.2 $\pm$ 0.4) $R_{d1}$; 
the corresponding values for the thin disc are 3.9$\pm$0.2 kms$^{-1}$ and (6.0$\pm$0.2) $R_{d2}$. The density averaged vertical velocity dispersion does not seem 
to reflect any component in particular, but remains constant at 6 kms$^{-1}$ at all $R$. The scaleheight from two component model and AGAMA seem to match fairly well both for the thin and thick disc cases. 
The multi-component stability parameter $Q_{N}$ in 3.6 $\mu$m has a minimum value of is 5.7, which implies that the disc is highly stable against 
axis-symmetric instabilities. In Figure 7, we present the posterior probability distribution and covariance plots of the parameters of the multi-component model of the
galactic disc of IC2233 with the stellar component modelled by $r$-band [Left Panel] and 3.6$\mu$m photometry [Right Panel]

\subsection{UGC00711}

The central vertical velocity dispersion in B-band is 18.4$\pm$ 0.9 kms$^{-1}$, and falls off exponentially with a scale length of (3.2$\pm$0.4) $R_{d1}$ where $R_{d1}$ is the
exponential scale length of the optical disk. We have calculated $Q_{N}$ as a function of R,we find that the minimum value is 4.5, 
indicating thedisc is highly stable. The central velocity dispersion in 3.6$\mu$m is 23.8$\pm$1.5 kms$^{-1}$, the dispersion falls off exponentially with 
scalelength (2.4$\pm$0.3)$R_{d1}$. The minimum value of $Q_{N}$ is 4.3. We note that the vertical velocity dispersion as well as the disk dynamical stability profiles 
of the optical and the 3.6  $\mu$m disc of UGC00711 match well well with each other, possibly confirming that they represent one and the same disc. 
In Figure 8, we present the posterior probability distribution and covariance plots of the parameters of the multi-component model of the galactic disc of UGC711 with the stellar component modelled by $r$-band [Left Panel] and 3.6$\mu$m photometry [Right Panel]

\end{document}